\documentclass[10pt,conference]{IEEEtran}

\usepackage{blindtext, graphicx}
\usepackage{color}
\usepackage[strings]{underscore}
\usepackage{algorithmicx}
\usepackage[ruled]{algorithm}
\usepackage{algpseudocode}
\usepackage{url}
\usepackage{listings}
\usepackage{amssymb,amsmath,amsthm}
\usepackage[compress,sort,space]{cite}
\usepackage{booktabs}
\usepackage{multirow}
\usepackage{comment}
\usepackage{pifont}
\usepackage[x11names]{xcolor}
\newcommand{\tool}{{{\sc{Alice}}}}

\newcommand{\ttt}[1]{\tt\small{#1}}

\usepackage{scrextend}
\usepackage{enumitem}
\usepackage{siunitx}

\definecolor{darkBlue}{rgb}{0.000000,0.000000,0.545098}
\definecolor{darkGreen}{rgb}{0.000000,0.392157,0.000000}
\definecolor{DarkGray}{gray}{0.4}
\definecolor{javared}{rgb}{0.6,0,0} 
\definecolor{javagreen}{rgb}{0.25,0.5,0.35} 
\definecolor{javapurple}{rgb}{0.5,0,0.35} 
\definecolor{javadocblue}{rgb}{0.25,0.35,0.75} 
\definecolor{lightgray}{gray}{0.95}
\definecolor{shadecolor}{RGB}{150,150,150}
\definecolor{blueA}{RGB}{204,229,255}
\definecolor{redA}{RGB}{112,0, 0}

\lstdefinestyle{MyJavaSmallStyle} {
  language=Java,
  frame=none,
  xleftmargin=15pt,
  stepnumber=1, 
  numbers=left, 
  numbersep=5pt,
  numberstyle=\tiny\color[gray]{0.777}, 
  belowcaptionskip=\bigskipamount,
  captionpos=b, 
  escapeinside={*'}{'*},
  tabsize=5,
  emphstyle={\bf},
  basicstyle=\scriptsize\ttfamily,
  keywordstyle=\color{javapurple}\bfseries,
  stringstyle=\color{javared},
  commentstyle=\color{javagreen},
  morecomment=[s][\color{javadocblue}]{/**}{*/},
  showspaces=false,
  columns=flexible,
  showstringspaces=false,
  morecomment=[l]{//},
  tabsize=2,
  breaklines=true
}

\algnewcommand\algorithmicforeach{\textbf{for each}}
\algdef{S}[FOR]{ForEach}[1]{\algorithmicforeach\ #1\ \algorithmicdo}

\makeatletter
\newcommand*{\rom}[1]{\expandafter\@slowromancap\romannumeral #1@}

\usepackage{fixltx2e}
\usepackage[margin=2cm]{geometry}
\usepackage[labelfont=bf]{caption}

\definecolor{mywhite}{RGB}{255,255,255}
\definecolor{mygray}{RGB}{220,220,220}
\definecolor{olivegreen}{RGB}{0,100,0}

\DeclareMathOperator*{\argmax}{argmax}
\newcommand{\coloneq}{\plimpl}
\newcommand{\plimpl}{~\operatorname{:-}~}

\newif\ifproofread

\newcommand{\changemarker}[1]{%
\ifproofread
\textcolor{blue}{#1}%
\else
#1%
\fi
}
\proofreadfalse
\definecolor{dkgreen}{rgb}{0,0.6,0}
\definecolor{gray}{rgb}{0.5,0.5,0.5}
\definecolor{mauve}{rgb}{0.58,0,0.82}
\definecolor{orangep}{rgb}{0.71, 0.43, 0.89}
\definecolor{orp}{rgb}{1, 0.7, 0.278}

\lstdefinelanguage{Scala}{
  keywords={typeof, new, true, false, catch,def,val, function, return, null, catch, switch, var, if, in, while, do, else, case, break, assert, declare, const, define,fun, ite, not, check,sat,String, Int},
  keywordstyle=\color{blue}\bfseries,
  ndkeywords={class, export,extends, boolean, throw, implements, import, Enumeration, ZipEntry,this, abstract, reduce, filter, map, reduceByKey, join},
  ndkeywordstyle=\color{mauve}\bfseries,
  otherkeywords={+, =>,<=, ==, >,< , ||},
  identifierstyle=\color{black},
  sensitive=false,
  comment=[l]{//},
  morecomment=[s]{/*}{*/},
  commentstyle=\color{purple}\ttfamily,
  stringstyle=\color{red}\ttfamily,
  morestring=[b]',
  morestring=[b]"
}

\theoremstyle{definition}

\begin{document}

\title{Active Inductive Logic Programming \\ for Code Search}

\author{Aishwarya Sivaraman, Tianyi Zhang, Guy Van den Broeck, Miryung Kim \\ University of California, Los Angeles \\
\{dcssiva, tianyi.zhang, guyvdb, miryung\}@cs.ucla.edu}

\markboth{Journal of \LaTeX\ Class Files,~Vol.~14, No.~8, August~2015}%
{Shell \MakeLowercase{\textit{et al.}}: Bare Advanced Demo of IEEEtran.cls for IEEE Computer Society Journals}

\IEEEtitleabstractindextext{%
\begin{abstract}
Modern search techniques either cannot efficiently incorporate human feedback to refine search results or cannot express structural or semantic properties of desired code. The key insight of our interactive code search technique {\tool} is that user feedback can be actively incorporated to allow users to easily express and refine search queries. We design a query language to model the structure and semantics of code as logic facts. Given a code example with user annotations, {\tool} automatically extracts a logic query from code features that are tagged as important. Users can refine the search query by labeling one or more examples as desired (positive) or irrelevant (negative). {\tool} then infers a new logic query that separates positive examples from negative examples via active inductive logic programming. 
Our comprehensive simulation experiment shows that {\tool} removes a large number of false positives quickly by actively incorporating user feedback. Its search algorithm is also robust to user labeling mistakes. Our choice of leveraging both positive and negative examples and using nested program structure as an inductive bias is effective in refining search queries. Compared with an existing interactive code search technique, {\tool} does not require a user to manually construct a search pattern and yet achieves comparable precision and recall with much fewer search iterations. A case study with real developers shows that {\tool} is easy to use and helps express complex code patterns.
\end{abstract}

}

\maketitle

\IEEEdisplaynontitleabstractindextext

%
\IEEEpeerreviewmaketitle

\section{Introduction}
\label{sec:introduction}
Software developers and tools often search for code to perform bug fixes, optimization, refactoring, etc. For example, when fixing a bug in one code location, developers often search for other similar locations to fix the same bug~\cite{kim2006memories, nguyen2010recurring, park2012empirical}. Text-based search techniques allow users to express search intent using keywords or regular expressions. However, it is not easy to express program structures or semantic properties using text-based search, thus hindering its capability to accurately locate desired code locations. TXL~\cite{cordy2006txl} and Wang et al.~\cite{wang2010matching} provide a domain-specific language (DSL) for describing structural and semantic properties of code. However, learning a new DSL for code search can be cumbersome and time consuming. 

Several techniques infer an underlying code search pattern from a user-provided example~\cite{andersen2010generic,pham2010detection,nguyen2010recurring, meng2011systematic}. These techniques adopt fixed heuristics to generalize a concrete example to a search pattern, which may not capture various search intent or allow a user to refine the inferred pattern. Lase~\cite{meng2013lase} and Refazer~\cite{rolim2017learning} use multiple examples instead of a single example to better infer the search intent of a user. However, this requirement poses a major usability limitation: a user must come up with multiple examples a priori. Critics allows a user to construct an AST-based search pattern from a single example through manual code selection, customization, and parameterization~\cite{zhang2015interactive}. However, users of Critics report that, the internal representation of a search pattern is not easy to comprehend and that they could benefit from some hints to guide the customization process. 

We propose an interactive code search technique, {\tool} that infers a search pattern by efficiently incorporating user feedback via active learning. {\tool} has three major components: 
(1) a novel user interface that allows a user to formulate an initial search pattern by tagging important features and iteratively refining it by labeling positive and negative examples, (2) a query language that models structural and semantic properties of a program as logic facts and expresses a search pattern as a logic query, and (3) an active learning algorithm that specializes a search pattern based on the positive and negative examples labeled by the user. In this interface, a user can start by selecting a code block in a method as a seed and by tagging important code features that must be included in the pattern. {\tool} then automatically lifts a logic query from the tagged features and matches it against the logic facts extracted from the entire codebase. Our query language models properties including loops, method calls, exception handling, referenced types, containment structures, and sequential ordering in a program. Therefore, our query language can easily express a search pattern such as `\textit{find all code examples that call the {\ttt readNextLine} method in a loop and handle an exception of type {\ttt FileNotFoundException}},' which cannot be accurately expressed by text-based approaches. 

Our active learning algorithm utilizes Inductive Logic Programming (ILP)~\cite{muggleton1994inductive,DBLP:ilpbook} to refine the search pattern. ILP provides an easy framework to express background knowledge as rules and is well-suited for learning from structured data. ILP does not assume a {\em flat} data representation and allows to capture structural properties of code not easily captured by other methods such as neural networks~\cite{neuralnetworks}. Additionally, ILP has been proven successful in program synthesis~\cite{muggleton1990efficient}, specification mining~\cite{cohen1994recovering, bratko1993inductive}, and model checking~\cite{alrajeh2013elaborating}. Since users can only inspect a handful of search results due to limited time and attention~\cite{starke2009working}, {\tool} allows the user to provide {\em partial feedback} by only labeling a few examples and gradually refine the search pattern in multiple search iterations. {\tool} uses an inductive bias to specialize the previous search pattern to separate the labeled positive and negative examples in each iteration. This feedback loop iterates until the user is satisfied with the returned result. 

We evaluate {\tool} using two benchmarks from prior work~\cite{meng2013lase,ahmad2018automatically}. These benchmarks consist of 20 groups of similar code fragments in large-scale projects such as Eclipse JDT. {\tool} learns an intended search pattern and identifies similar locations with 93\% precision and 96\% recall in three iterations on average, when the user initially tags only two features and labels three examples in each iteration.

A comprehensive simulation experiment shows that labeling both positive and negative examples are necessary to obtain the best search results, compared to labeling only positives or only negatives. It is because negative examples quickly reduce the search space, while positive examples are used to capture the search intent. 
We vary the number of labeled examples and find that labeling more examples in each iteration does not necessarily improve the final search accuracy. This indicates that a user can label only a few in each iteration to reach good accuracy eventually. {\tool} thus alleviates the burden of labeling many examples at once, as its active learning can leverage partial feedback effectively. The comparison with an existing technique Critics~\cite{zhang2015interactive} shows that {\tool} achieves the same or better accuracy with fewer iterations. 
\begin{figure}
\includegraphics[width=\linewidth]{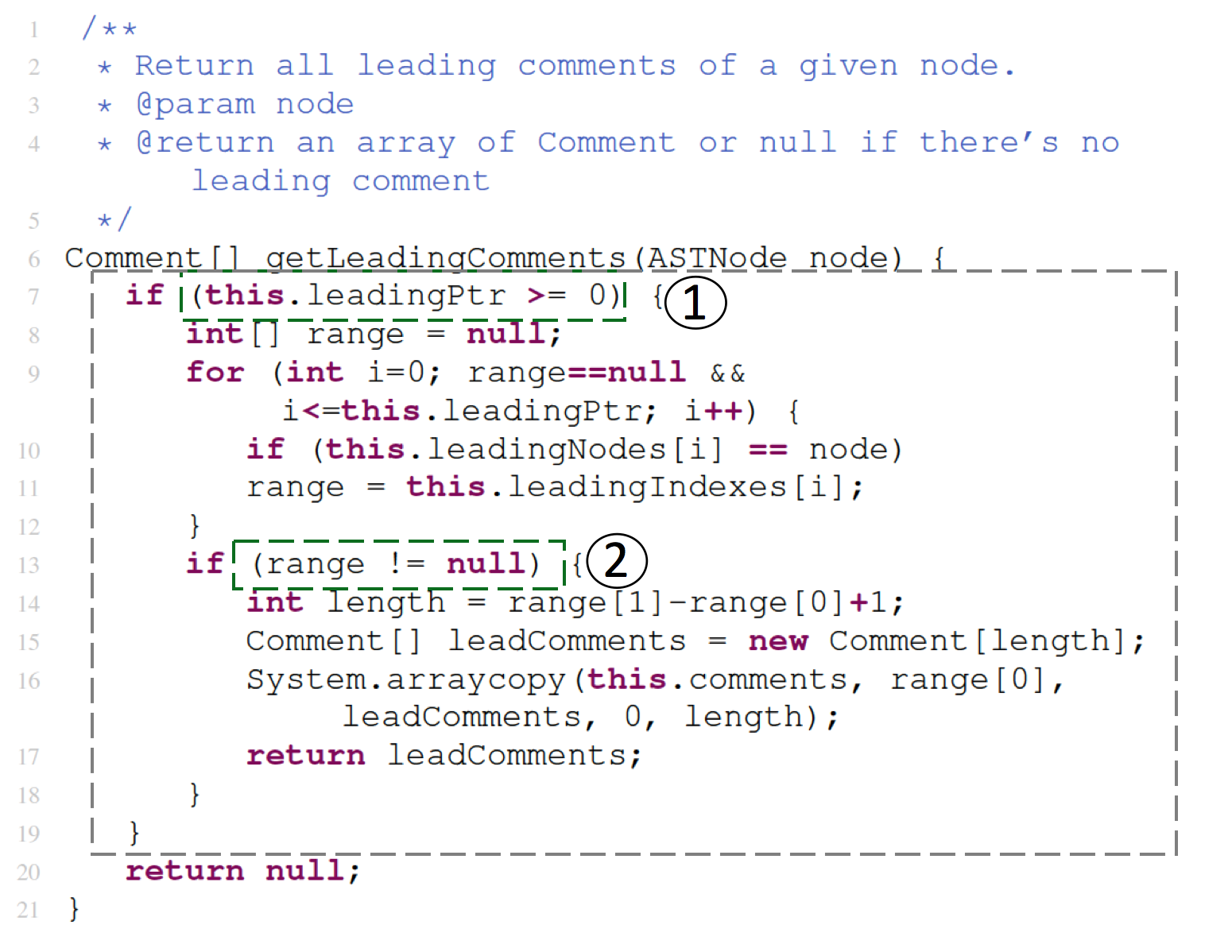}
\caption{Bob tags {\ttt if} condition {\ttt range!=null} and {\ttt if} condition {\ttt this.leadingPtr>=0} as must-have code elements.}
  \label{fig:sel2}
  \vspace{-5mm}
\end{figure}

This human-in-the-loop approach in {\tool} opens a door for incorporating user feedback in a variety of software engineering techniques, e.g., bug detection, code optimization, refactoring, etc. For example, a code optimization framework, Casper, searches for Java programs with loops that sequentially iterate over data and translates the loops to semantically equivalent {\ttt map-reduce} programs~\cite{ahmad2016leveraging}. However, the search patterns in Casper are hardcoded and cannot be easily expressed by a user. {\tool} can aid in such scenarios by allowing a user to interactively construct a search pattern.
\changemarker{Our paper makes the following contributions.\begin{itemize}
    \item We present a novel approach called {\tool} that integrates active learning and inductive logic programming to incorporate partial user feedback and refine code search patterns. {\tool} is instantiated as an Eclipse plug-in and the tool is available online.\footnote{Our tool and dataset are available at\\https://github.com/AishwaryaSivaraman/ALICE-ILP-for-Code-Search} 
    \item We conduct a comprehensive simulation experiment that investigates the effectiveness of {\tool} using different inductive biases, different numbers of labeled examples, and different numbers of annotated code features. An additional robustness experiment shows that our search algorithm is resilient to labeling mistakes by flagging contradictory examples labeled by a user.  
    \item We conduct a case study with real developers, demonstrating that participants can easily interact with {\tool} to search desired code by simply inspecting and labeling code examples. On average, each participant labels two examples in each search iteration and spends about 35 seconds on each example. 
\end{itemize}}

\section{Motivating Example}
\label{sec:motive}
\begin{figure}
\begin{lstlisting}[style=MyJavaSmallStyle]
public int getExtendedStartPosition(ASTNode node) {
    if (this.leadingPtr >= 0) {
        int[] range = null;
        for (int i=0; i<=this.leadingPtr; i++) {
            if (this.leadingNodes[i] == node) 
            range = this.leadingIndexes[i];
        }        
        if (range != null) {
            return  
            this.comments[range[0]].getStartPosition() ;
        }
    }
    return node.getStartPosition();
}
\end{lstlisting}
\caption{Bob labels this example as positive. This example is syntactically similar to Figure 1, but has a different {\ttt loop} condition {\ttt i<=this.leadingPtr}.}
  \label{fig:posexample}
  \vspace{-3mm}
\end{figure}

This section describes the code search process using {\tool} with a real-world example drawn from Eclipse Java Development Toolkit (JDT). Eclipse JDT provides basic tools and library APIs to implement and extend Eclipse plug-ins. As an Eclipse JDT developer, Bob wants to update the {\ttt getLeadingComments} method (Figure~\ref{fig:sel2}) to return an empty {\ttt Comment} array when there are no leading comments, instead of returning {\ttt null}, which may cause a {\ttt NullPointerException} in a caller method. Before modifying the code, Bob wants to check other similar locations. 

Bob could use a text-based search tool to find other code fragments similar to the {\ttt getLeadingComments} method. For example, Bob could use {\ttt Comment} and {\ttt arrayCopy} as keywords to find code locations that reference the {\ttt Comment} type and call the {\ttt arrayCopy} method. Such text-based techniques are prone to return a large number of search results, many of which merely contain the same keywords but have significantly different code structures or irrelevant functionality. In the JDT codebase, searching with the {\ttt arrayCopy} keyword returns 1854 method locations. It is prohibitively time-consuming to inspect all of them. Prior work has shown that developers only inspect a handful of search results and return to their own code due to limited time and attention~\cite{brandt2009two, starke2009working, duala2012asking}. %

To narrow down the search results, it would be useful to describe the structural properties of {\ttt getLeadingComments}. For example, a code snippet must contain an {\ttt if} condition that checks for {\ttt >=0} and contain a loop inside the {\ttt if} block. Instead of requiring a user to prescribe such a structural pattern manually, {\tool} enables the user to tag important code features and label positive vs.~negative examples instead.

\begin{figure}
\begin{lstlisting}[style=MyJavaSmallStyle]
public void checkComment() {
	...
	if (lastComment >= 0) {
		this.modifiersSourceStart = this.scanner.commentStarts[0]; 
	
		while (lastComment >= 0 && this.scanner.commentStops[lastComment] < 0) lastComment--;
		if (lastComment >= 0 && this.javadocParser != null) {
			if (this.javadocParser.checkDeprecation(
					this.scanner.commentStarts[lastComment],
					this.scanner.commentStops[lastComment] - 1)) {
				checkAndSetModifiers(AccDeprecated);
			}
			this.javadoc = this.javadocParser.docComment;
		}
	}
}
\end{lstlisting}
\caption{Bob labels this example as negative. Though it has a similar {\ttt if} check {\ttt lastComment>=0}, its program structure is significantly different from Figure 1.}
  \label{fig:negexample}
  \vspace{-5mm}
\end{figure}

\noindent{\bf Iteration 1. Select a Seed Example and Annotate Important Features.} Bob first selects a code block of interest in the {\ttt getLeadingComments} method (lines 7-19 in Figure~\ref{fig:sel2}). He {\em annotates} an {\ttt if} condition, {\ttt range!=null} ({\large\color{black} \textcircled{\small 2}} in Figure~\ref{fig:sel2}), as a code element that must be included in the search pattern, since he posits that similar locations would have a {\ttt null} check on the range object. Such must-have code elements are colored in green. {\tool} then automatically constructs a search query based on the selected code and annotations. Since other developers may rename variables in different contexts, {\tool} generalizes the variable name {\ttt range} in the tagged {\ttt if} condition to match a {\ttt null} check on any variable. Among all 12633 Java methods in the Eclipse JDT codebase, {\tool} returns 1605 methods with a {\ttt null} check, which are too many to examine. Bob now has two choices. He can annotate more features or label a few examples to reduce the search results. Bob tags another {\ttt if} condition, {\ttt this.leadingPtr>=0}, as another must-have element ({\large \textcircled{\small 1}} in Figure~\ref{fig:sel2}). The field name {\ttt leadingPtr} is also generalized to match any variable name. {\tool} refines the previous query and returns ten methods that both perform a {\ttt null} check and contain an {\ttt if} condition checking, if a field is no less than 0. The query is shown below and its syntax is detailed in Section~\ref{sec:approach}.

{
\centering
\vspace{0.10cm}
\fbox{
\begin{minipage}[t][3.1em][t]{23.5em}
{\footnotesize{\ttt 
query(X) $\coloneq$ methoddec(X),
\mbox{~~}contains(X,IF\textsubscript{0}), iflike(IF\textsubscript{0},"this.*$>=$0"),
\mbox{~~}contains(X,IF\textsubscript{2}), iflike(IF\textsubscript{2},".*!=null").}} 
\end{minipage}}
\vspace{0.05cm}
}

\begin{figure}
      \includegraphics[width=\linewidth]{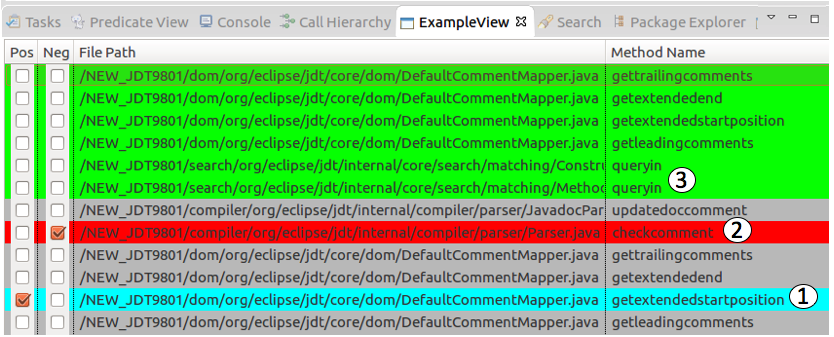}
  \caption{Bob labels some examples as desired and some as irrelevant. Green {\large \textcircled{\small 3}} indicates newly returned examples, Red {\large \textcircled{\small 2}} indicates a previously marked negative examples, and Cyan {\large \textcircled{\small 1}} indicates a previously marked positive example.}
  \label{fig:label}
  \vspace{-3mm}
\end{figure}

\noindent{\bf Iterations 2 to N. Active Learning by Labeling Positive and Negative Examples.} After inspecting two examples returned from the previous iteration, Bob labels one example (Figure~\ref{fig:posexample}) as positive and another one (Figure~\ref{fig:negexample}) as negative. 
Bob labels Figure~\ref{fig:negexample} as irrelevant, since this example has similar {\ttt if} conditions (line 3), but does not compute the range of a code comment or returns a range. 
By incorporating this feedback, {\tool} learns a new query that includes a unique structural characteristic---{\em the null check is contained by the if check for an index field greater or equal to 0}---that appears in both the initially selected example and the positive example but not in the negative example, as shown below. As a result, {\tool} narrows down to six examples ({\large \textcircled{\small 3}} in Figure~\ref{fig:label}). A user can further refine the search query and continue to iterate by labeling examples. As {\tool} uses a top-down search algorithm to {\em specialize} its search results in each iteration, the resulting set is always a subset of the previous results. 

{
\centering
\vspace{0.10cm}
\fbox{
\begin{minipage}[t][3.1em][t]{23.5em}
{\footnotesize{\ttt 
query(X) $\coloneq$ methoddec(X),
\mbox{~~}contains(X,IF\textsubscript{0}), iflike(IF\textsubscript{0},"this.*$>=$0"),
\mbox{~~}contains(IF\textsubscript{0},IF\textsubscript{2}), iflike(IF\textsubscript{2},".*!=null").}} 
\end{minipage}}}

\section{Approach}
\vspace{-0.1cm}


\subsection{Machine Learning Approach}
Traditional code search techniques retrieve code examples by matching user-specified patterns. In contrast, {\tool}'s central thesis is that such patterns need not be fully specified by the user, and can instead be induced from specific user interactions. This is fundamentally a machine learning approach to code search: the intended pattern is a hypothesis that is learned from data.
To achieve this goal, {\tool} integrates several learning paradigms.

First, it requires labels that categorize code examples as either matching the intended hypothesis or not.
Classical supervised learning would require the user to provide prohibitively many labels. {\tool}, therefore, employs \emph{active learning}~\cite{settles2012active}, where the learner has the ability to present some unlabeled examples to the user during learning and ask to provide a label.
In a well-designed active learning setup, the user only needs to provide a minimal number of labels for the learner to reliably find the correct hypothesis.

\changemarker{Second, most machine learning approaches represent data as feature vectors, which cannot easily express the structure of source code.}
\emph{Inductive logic programming}~(ILP)~\cite{muggleton1994inductive,de2008logical} is a form of relational learning that supports structured data encoded as logical facts.
In particular, {\tool} uses the logical predicates listed in the first column of Table~\ref{tab:predicates} to represent each code example in a factbase. The next section describes this process in~detail.

Given that the data is now in a suitable form, ILP  aims to learn a hypothesis by constructing a logical (Prolog~\cite{flach1994simply}) query that returns exactly those code IDs that the user is searching for. Queries in {\tool} are represented as definite clauses. For example, in the queries shown in Section~\ref{sec:motive}, we are looking for values of the logical variable $\ttt X$ for which the body of the query is true. The body is true if there exists a value for the remaining logical variables (uppercase), such that each atom in the body is also found in the factbase. 
The process of learning a query is centered around two operations: \emph{generalization} and \emph{specialization}. Generalization changes the query to return more results, e.g., removing atoms from the body, or replacing constants with variables. Specialization is the reverse operation, yielding a query with fewer results.
An ILP learner uses these operations to search the space of all hypotheses for the right one.

Finally, even with active learning, {\tool} operates in a regime where little labeled data is available. We know from learning theory that this is only possible when the learner has a strong \emph{inductive bias}. That is, the learner already incorporates a lot of knowledge about the learning task, before even seeing the data.
We address this in three ways. 
First, ILP affords the specification of declarative background knowledge, which helps to create a more meaningful inductive bias as it applies to code search. Table~\ref{tab:background} shows {\tool}'s background knowledge, which enables {\tool} to use additional predicates to construct richer queries.
Second, we allow the user to annotate important code features in order to obtain a stronger inductive bias. 
Third, {\tool} adopts a specialization procedure that is specifically designed to capture different program structures to strengthen the inductive bias. \changemarker{We implement our own realtime ILP system based on the high-performance YAP Prolog engine~\cite{costa2012yap}.}

\subsection{Logic Fact Extraction} 
{\tool} creates a factbase from a given code repository using the predicates described in Table~\ref{tab:predicates}. It parses program files to Abstract Syntax Trees (ASTs) and traverses the ASTs to extract logic facts from each method. Predicates {\ttt if} and {\ttt loop} capture the control flow within a method; {\ttt methodcall} represents the relationship between a method and its caller; {\ttt exception} captures the type of handled exceptions; {\ttt type} captures the type of defined and used variables; {\ttt contains} describes the relationship between each AST node and its parent recursively; {\ttt before} captures the sequential ordering of AST nodes. 
Specifically, {\ttt before(id$_1$,id$_2$)} is true, when node {\ttt id$_1$} comes before node {\ttt id$_2$} in pre-order traversal while excluding any direct or transitive containment relationship. For Figure~\ref{fig:sel2}, {\ttt loop(loop\textsubscript{1}, "range==null \&\& i$<=$this.leadingPtr")} comes before {\ttt if(if\textsubscript{2}, "range!=null")} creating {\ttt before(loop\textsubscript{1}, if\textsubscript{2})}. Ground facts are then loaded into a YAP Prolog engine~\cite{costa2012yap} for querying. 

\subsection{Generalization with Feature Annotation (Iteration 1)} 
When a user selects a code block of interest, {\tool} generates a specific query which is a conjunction of atoms constructed using the predicates in Table~\ref{tab:predicates}. Each atom in the query is grounded with constants for its location {\ttt ID} and value representing the AST node content. {\tool} replaces all {\ttt ID} constants in the query with logical variables to generate an initial candidate hypothesis $h_{0}$. For example, one ground atom in the query is of the form {\ttt if(if,"range!=null")} and its variablized atom is of the form {\ttt if(IF,"range!=null")}.
To find more examples, {\tool} generalizes the hypothesis by dropping atoms in $h_0$ other than the user annotated ones, producing $h_1$. 

\begin{table}
\centering
\caption{Predicates in Logical Representation}
\label{tab:predicates}
\begin{tabular}{@{}ll@{}}
\toprule
Fact Predicates & Rule Predicates\\ \midrule
$\mathtt{if(ID,COND)}$ & $\mathtt{iflike(ID,REGEX)}$\\
$\mathtt{loop(ID,COND)}$ & $\mathtt{looplike(ID,REGEX)}$\\
$\mathtt{parent(ID,ID)}$ & $\mathtt{contains(ID,ID)}$\\
$\mathtt{next(ID,ID)}$ & $\mathtt{before(ID,ID)}$\\
$\mathtt{methodcall(ID,CALL)}$ & \\
$\mathtt{type(ID,NAME)}$ & \\
$\mathtt{exception(ID,NAME)}$ & \\
$\mathtt{methoddec(ID)}$ & \\
\bottomrule
\end{tabular}
\end{table}

\begin{table}
\centering
\caption{Background Knowledge}
\label{tab:background}
\begin{tabular}{@{}rcl@{}}
\toprule
\multicolumn{3}{c}{Prolog Rules}\\ 
\midrule
$\mathtt{iflike(ID,REGEX)}$ & $\coloneq$ 
    & $\mathtt{if(ID,COND)},$\\
  & & $\mathtt{regex\_match(COND,REGEX)}.$\\
$\mathtt{looplike(ID,REGEX)}$ & $\coloneq$ 
    & $\mathtt{loop(ID,COND)},$\\
 &  & $\mathtt{ regex\_match(COND,REGEX)}.$\\
$\mathtt{contains(ID_1,ID_2)}$ & $\coloneq$ 
    & $\mathtt{parent(ID_1,ID_2)}.$\\
$\mathtt{contains(ID_1,ID_3)}$ & $\coloneq$ 
    & $\mathtt{parent(ID_1,ID_2)},$\\
 &  & $\mathtt{contains(ID_2,ID_3)}.$\\
$\mathtt{before(ID_1,ID_2)}$ & $\coloneq$ 
    & $\mathtt{next(ID_1,ID_2)}.$\\
$\mathtt{before(ID_1,ID_3)}$ & $\coloneq$ 
    & $\mathtt{next(ID_1,ID_2)},$\\
 &  & $\mathtt{ before(ID_2,ID_3)}.$\\
\bottomrule
\end{tabular}
\vspace{-3mm}
\end{table}


\noindent \textbf{Regex Conversion.} {\tool} further abstracts variable names in $h_1$ to identify other locations that are similar but have different variable names. {\tool} converts predicates {\ttt if} and {\ttt loop} to {\ttt iflike} and {\ttt looplike} respectively. As defined in Table~\ref{tab:background}, {\ttt iflike} and {\ttt looplike} are background rules that match a regular expression ({\ttt REGEX}) with ground conditions ({\ttt COND}) in the factbase. For instance, each variable name in {\ttt loop(ID,"range==null \&\& i$<=$ this.leadingPtr")} is converted from a string constant to a Kleene closure expression, generating {\ttt looplike(ID,".*==null \&\& .*$<=$this.*")}. The output of this phase is a generalized query $h_{2}$ and a set of code examples that satisfy this query. This example set is displayed in the \emph{Example View} in Figure~\ref{fig:label}.

\subsection{Specialization via Active Learning (Iterations 2 to N)} 
In each subsequent search iteration, given the set of positive examples ($P$) and the set of negative examples ($N$) labeled by a user, the previous hypothesis $h_{i-1}$, which is a conjunction of atoms, is {\em specialized} to $h_{i}$ by adding another atom to exclude all negatives while maximally covering positives. The specialization function is defined below.
\begin{align*}
     {\ttt Specialize}(h_{i-1},P,N) = \argmax_{h_i} \sum_{p \in P}[p \models h_i] \\
     \text{ such that } h_i \models h_{i-1} \text{ and } \forall n \in N, n \not\models h_i
\end{align*}
Suppose all positive examples call \texttt{foo} and all negative examples call \texttt{bar} instead of \texttt{foo}. We add a new atom \texttt{calls(m,"foo")} to specialize $h_{i-1}$, which distinguishes the positives from the negatives. As a design choice, our active ILP algorithm is {\em consistent} (i.e., not covering any negative example) but is not {\em complete} (i.e., maximally satisfying positive examples). Our learning algorithm is monotonic in that it keeps adding a new conjunctive atom in each search iteration. 
This specialization procedure always returns a subset of the previous search results obtained by $h_{i-1}$. This feedback loop continues to the $n$-th iteration until the user is satisfied with the search results.

Given the large number of candidate atoms, \emph{inductive bias} is required to guide the specialization process of picking a discriminatory atom. {\tool} implements three inductive biases, which are described below. The effectiveness of each bias is empirically evaluated in Section~\ref{sec:evaluation}.

\begin{itemize}[leftmargin=*]

\item{\textbf{Feature Vector.}} This bias considers each code block to have a {\em flat} structure. The feature vector bias does not consider the nested structure or sequential code order. It specializes by adding a random atom that reflects the existence of loops, if checks, method calls, types, or exceptions in the code block. 
It is used as the baseline bias in the evaluation since it does not utilize any structural information such as containment and ordering.

\item{\textbf{Nested Structure.}} This bias utilizes the containment structure of the seed code example to add atoms. In addition to adding an atom corresponding to the AST node, the bias adds a {\ttt contains} predicate to connect the newly added atom to one that already appears in the query. Consider an AST with root node $A$, whose children are $B$ and $C$; $B$ has children $D$ and $E$, and $C$ has child $F$. Suppose that $h$ includes an atom referring to $B$. Then based on the containment relationships, {\tool} selects one of $D$ or $E$ to specialize the query for the next iteration, not $F$. If there are no available children, or if this query fails to separate positives from negatives, it falls back to the parent of $B$ or its further ancestors to construct the next query. We choose this nested structure bias as default since it empirically achieves the best performance~(detailed in Section~\ref{eval:bias}).

\item{\textbf{Sequential Code Order.}} This bias uses sequential ordering of code in the seed example to determine which atom to add next. Consider an example AST with root node $A$ and children $B$ and $C$; $C$ itself has children $D$ and $E$. Atoms {\ttt before($B$,$D$)}, {\ttt before($B$,$C$)}, and {\ttt before($B$,$E$)} are generated according to the rules in Table~\ref{tab:background}. Given a query that contains atoms referring to $B$, {\tool} now chooses one of $C$, $D$, or $E$, to connect to $B$ using the {\ttt before} predicate, and adds this node to the query. If there are no available atoms to add, or if this query fails to separate positives from negatives, it falls back to the original feature vector bias. 
\end{itemize} 

An alternative approach is to carry out logical generalization where we generate a query by generalizing positive examples and taking the conjunction with the negation of negative example generalization. As a design choice, we do not allow negations in our query for realtime performance, since supporting negations would significantly increase the search space and execution time. 
 \label{sec:approach}


\section{Simulation Experiment}
\label{sec:evaluation}

\begin{table*}[h]
\centering
\caption{Simulation Experiment Dataset and Results Summary}
\label{tab:dataset}
\resizebox{\textwidth}{!}{
\begin{tabular}{@{}crrrrrrrcrr@{}}
\toprule
 \multicolumn{8}{c}{\sf{Groups}} & \multirow{2}{*}{{\sf{Repo Revision}}} & \multirow{2}{*}{\sf{\#Methods}} & \multirow{2}{*}{\sf{Factbase Size}}\\
   \cmidrule{1-8}
\sf{ID} & \sf{LOC} & \sf{Snippets} & \sf{Precision} & \sf{Recall} & \sf{F1} & \sf{\#Iterations} & \sf{Query Length} & & & \\
\midrule
 1  & 14 & 4 & 1.0 & 0.75 & 0.86 & 3 & 12 &JDT 9801 & 12633 & 541360\\
\cmidrule{1-3}
 2 & 44 & 9 & 1.0 & 0.91 & 0.95 & 3 & 11&\multirow{2}{*}{JDT 10610} & \multirow{2}{*}{13165} & \multirow{2}{*}{570049}\\
 3 & 648 & 5 & 1.0 & 0.92 & 0.96 & 2 & 11&\\
\cmidrule{1-3} 4 & 8 & 6 & 1.0 & 1.0 & 1.0 & 4 & 18 &
\multirow{3}{*}{SWT 16739} & \multirow{3}{*}{35863} & \multirow{3}{*}{918818} \\
 5 & 47 & 3 & 1.0 & 1.0 & 1.0 & 4 & 13&\\
 6 & 11 & 7 & 1.0 & 1.0 & 1.0 & 2 & 12&\\
\cmidrule{1-3}
 7 & 19 & 3 & 1.0 & 1.0 & 1.0 & 3 & 11&
 \multirow{8}{*}{SWT 3213515} & \multirow{8}{*}{20899}  & \multirow{8}{*}{663980}\\
 8 & 3  & 6 & 1.0 & 1.0 & 1.0 & 5 &18&\\
 9 & 18 & 5 & 1.0 & 1.0 & 1.0 & 3 &11&\\
 10 & 28 & 10 & 1.0 & 1.0 & 1.0 & 4 &13&\\
 11 & 112 & 4 & 1.0 & 1.0 & 1.0 & 3 &17&\\
 12 & 30 & 3 & 1.0 & 1.0 & 1.0 & 5 &11&\\
 13 & 34 & 3 & 1.0 & 1.0 & 1.0 & 3 &15&\\
 14 & 18 & 7 & 1.0 & 1.0 & 1.0 & 3 &12&\\
 \midrule
 \textbf{Average} & 74& 5& 1.0 & 0.97 &  0.98 & 3.4 & 13& & 20640&673551 \\
 \midrule
  15 & 10 & 11& 1.0 & 1.0 & 1.0 & 1 & 5 & Arith& 31 &316\\
 \cmidrule{1-3}
  16 & 16 & 6 & 1.0 & 1.0 & 1.0 & 1 &5 &Big$\lambda$& 15 & 498\\
 \cmidrule{1-3}
  17 & 9 & 6 & 1.0 & 0.66 & 0.8 & 1 &3&\multirow{2}{*}{Phoenix} & \multirow{2}{*}{30} & \multirow{2}{*}{783} \\
 18 &  37& 2& 1.0 & 1.0 & 1.0 & 2 & 5&\\
 \cmidrule{1-3}
  19 & 6 & 3 & 0.3 & 1.0 & 0.46 & 1 &3 &\multirow{2}{*}{Stats} & \multirow{2}{*}{60} & \multirow{2}{*}{1343} \\
  20 & 9 & 2 & 0.2 & 1.0 & 0.33 & 1 & 3 & \\
\midrule
\textbf{Average} & 15& 5& 0.75 & 0.94&  0.76 & 1.1 & 4& & 34& 735 \\
\midrule
\textbf{Total Average}  &56 & 5& 0.93 & 0.96&  0.92 & 2.7 &11 & & 10337& 337143\\
\bottomrule
\end{tabular} 
}%
\vspace{-3mm}
\end{table*}

We systematically evaluate the accuracy and effectiveness of {\tool} by assessing different search strategies and by simulating various user behaviors.

\noindent{\bf Dataset.} We use two complementary sets of similar code fragments as the ground truth for evaluation, as shown in Table~\ref{tab:dataset}. The first dataset is drawn from the evaluation data set of LASE~\cite{meng2013lase}. This dataset consists of groups of syntactically similar code locations in Eclipse JDT and SWT, where developers need to apply similar bug fixes. We select groups with more than two similar locations, resulting in 14 groups in total. Each group contains an average of five similar code fragments and each fragment contains a median of 24 lines of code, ranging from 3 to 648. \changemarker{{\tool} extracts an average of 670K logic facts from each repository.} The second data set is from  the evaluation dataset of Casper~\cite{ahmad2018automatically}, an automated code optimization technique. This dataset consists of groups of similar code fragments that follow the same data access patterns (e.g., a sequential loop over lists) and can be systematically optimized by Casper. 
By evaluating {\tool} with both datasets, we demonstrate that {\tool} is capable of accurately and effectively searching code in two different usage scenarios---bug fixing and code optimization. Because the second data set includes only the relevant files not the entire codebase, we cannot reliably assess the rate of false positives. Hence we exclude this second dataset when assessing the impact of individual biases, annotations, and labeling effort in Sections~\ref{eval:bias},~\ref{eval:bothlabel},~\ref{eval:features}, and~\ref{eval:lables}.   

\noindent{\bf Experiment environment.} All experiments are conducted on a single machine with an Intel Core i7-7500U CPU (2.7GHz, 2 cores/4 threads, x64 4.13.0-31-generic), 16GB RAM, and Ubuntu 16.04 LTS. We use YAP Prolog (version 6.3.3), a high-performance Prolog engine \cite{costa2012yap} to evaluate search queries. 

We write a simulation script to randomly select a code fragment in each group as the seed example. In the first iteration, the script randomly tags $k$ important features that represent control structures, method calls, and types in the seed example. In each subsequent iteration, it simulates the user behavior by randomly labeling $n$ examples returned in the previous iteration. If a code example appears in the ground truth, it is labeled as positive. Otherwise, it is labeled as negative. The script terminates if no more examples can be labeled in a new search iteration. To mitigate the impact of random choices, we repeat the simulation ten times and report the average numbers for each group. 

\noindent{\bf Result summary.} Table~\ref{tab:dataset} summarizes the precision, recall, and F1 score of {\tool} in the final search iteration.
When setting $k$ and $n$ to two and three respectively, {\tool} empirically achieves the best result, which we detail in Sections~\ref{eval:features} and~\ref{eval:lables}. On average, {\tool} successfully identifies similar code locations with 93\% precision and 96\% recall in 2.7 search iterations. {\tool} achieves 100\% precision and 97\% recall in the first dataset, while it achieves 75\% precision and 94\% recall in the second dataset. The reason is that the second dataset contains code fragments that loop over a double array with no write to output operations, which is a semantic constraint imposed by Casper~\cite{ahmad2018automatically} for loop optimization. However, {\tool} does not learn predicates that differentiate read and write operations on an array and therefore returns a large number of code fragments that write double arrays in a loop, which leads to low precision. 

In the first search iteration, \changemarker{138 methods} are returned by {\tool} on average (median: 23 and maximum: 2352). This set is large, motivating our approach to leverage partial feedback (i.e., only three labeled examples at a time) as opposed to labeling all returned results at once. {\tool} effectively incorporates partial feedback to reduce the number of returned examples by 93\% in the second iteration. \changemarker{Since {\tool} adopts a top-down search process and always starts with a general query, the initial search result tends to include all code locations in the ground truth. Therefore, the first search iteration often starts with 100\% recall and relatively low precision. The precision then improves gradually, as the query is specialized to eliminate false positives in the search results.}  
\changemarker{In Table~\ref{tab:dataset}, column query length represents the number of atoms in the final query. On average the query length inferred by {\tool} contains 11 atoms.}
\vspace{-1mm}
\subsection{Impact of Inductive Bias} \label{eval:bias}

During query specialization, an inductive bias is used to effectively navigate the space of all hypotheses. We evaluate the effectiveness of each of the three implemented inductive biases discussed in Section~\ref{sec:approach}: (1) feature vector, (2) nested structure, and (3) sequential order. 

\begin{figure}[htb]
      \includegraphics[width=\linewidth]{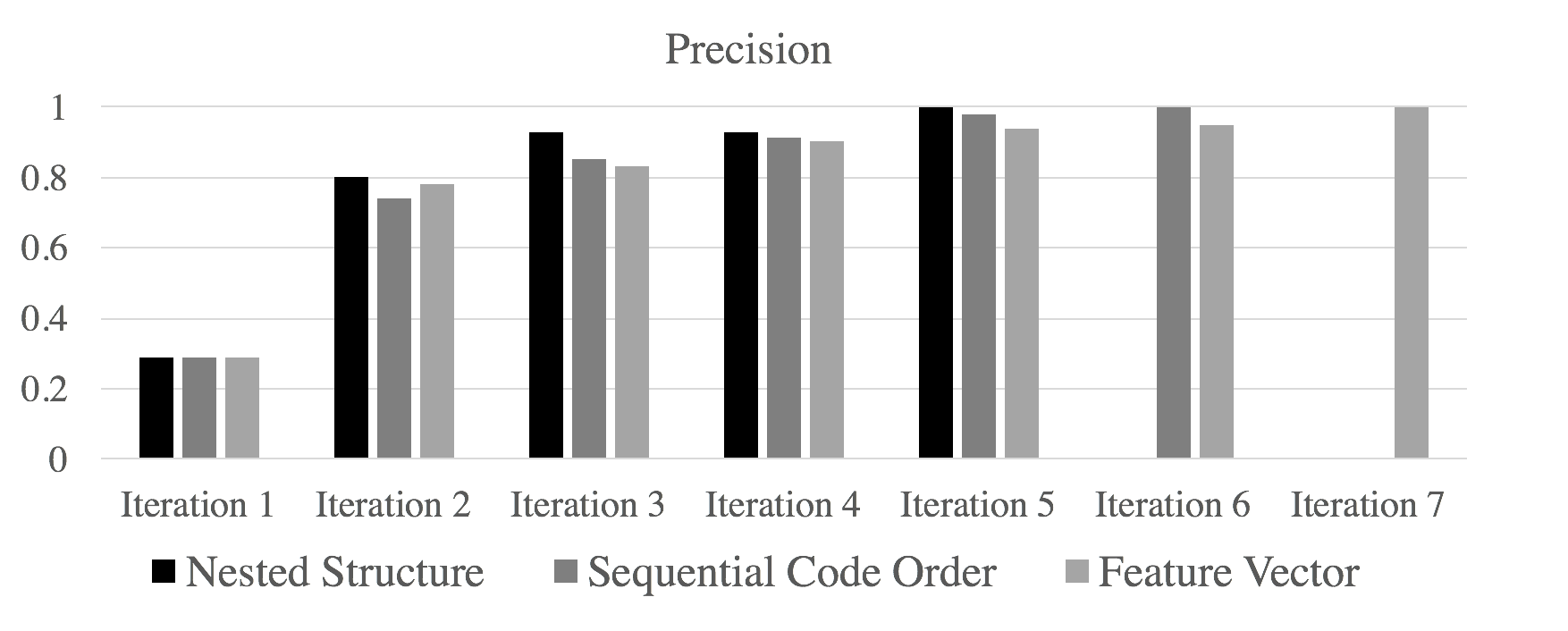}
  \caption{Precision of {\tool} using different inductive biases}
  \label{fig:biaspres}
\vspace{-3mm}
\end{figure}
\begin{figure}[htb]
      \includegraphics[width=\linewidth]{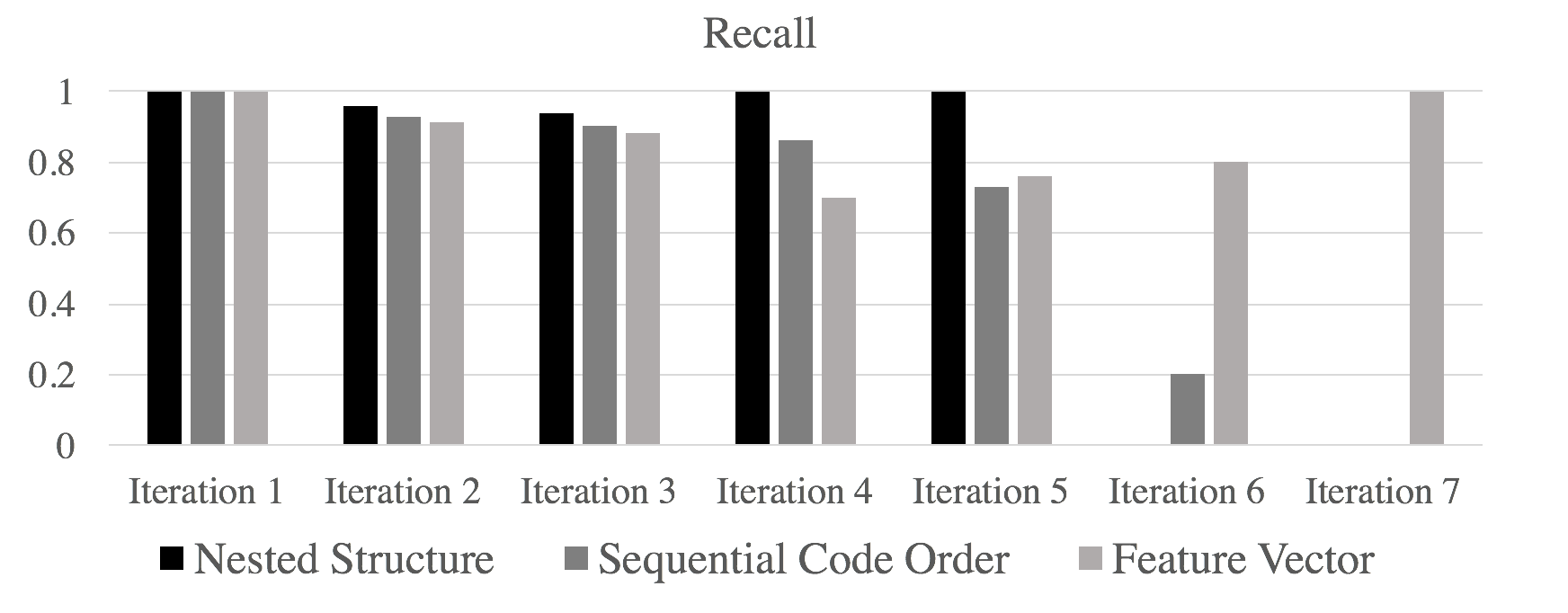}
  \caption{Recall of {\tool} using different inductive biases}
  \label{fig:biasf1}
\vspace{-3mm}
\end{figure}

Figures~\ref{fig:biaspres} and~\ref{fig:biasf1} show the effectiveness of each bias in terms of precision and recall, averaged across ten runs. Overall, the nested structure bias converges fast by taking fewer iterations to reach the highest F1 score. The sequential order bias performs better than the feature vector bias, converging in six iterations as opposed to seven. Although both the sequential order bias and the nested structure bias perform well in terms of precision, encountering the same sequential order between statements or methods is not common in the data set. Therefore, the sequential order bias has a lower recall. The nested structure bias adds atoms based on containment relationships and hence it filters out false positives early and converges faster. 

\vspace{-4mm}
\begin{figure}[ht]
	\begin{lstlisting}[style=MyJavaSmallStyle]
	checkWidget ();
	if (!parent.checkData (this, true)) error (SWT.ERROR_WIDGET_DISPOSED);
	return font != null ? font : parent.getFont ();
	\end{lstlisting}
	\caption{A code snippet from Group 8 in Table~\ref{tab:dataset}}
	\label{fig:evalbiasexample}
	\vspace{-3mm}
\end{figure}

\begin{figure}[!h]
      \includegraphics[width=\linewidth]{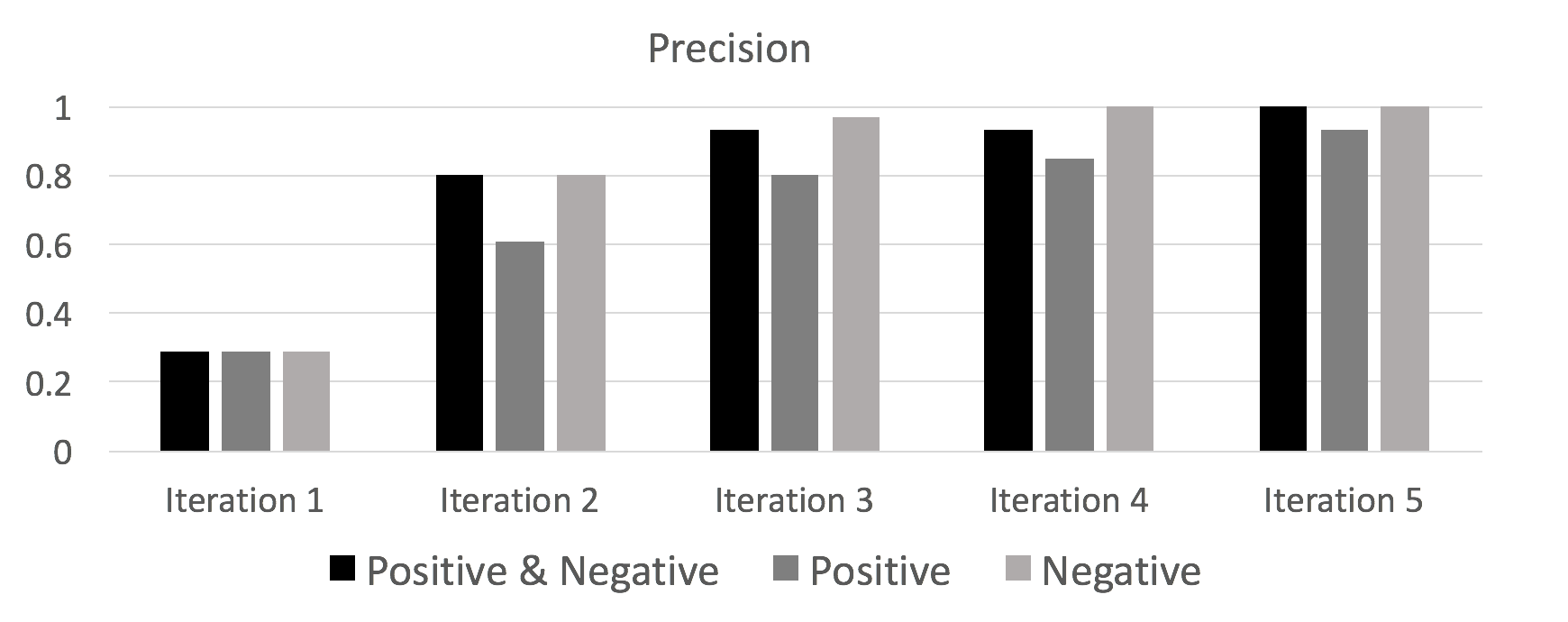}
  \caption{Precision with different types of labeled examples}
  \label{fig:onlyposnegpres}
  \vspace{-3mm}
\end{figure}

\begin{figure}[!h]
      \includegraphics[width=\linewidth,scale=0.1]{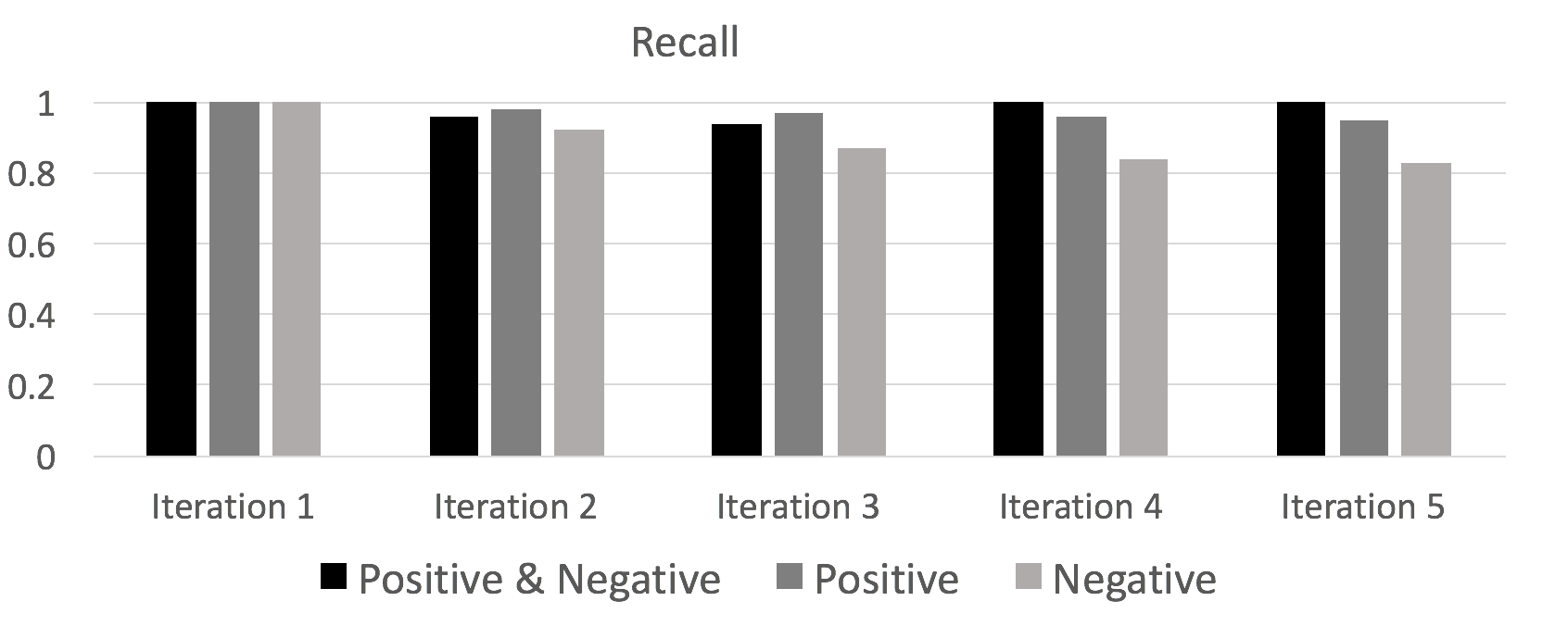}
  \caption{Recall with different types of labeled examples}
  \label{fig:onlyposnegpresrecall}
  \vspace{-3mm}
\end{figure}

Consider Figure~\ref{fig:evalbiasexample}, taken from group 8 in Table~\ref{tab:dataset}. The first iteration returns 90 examples when annotating \texttt{checkData} and \texttt{error}. The feature vector bias takes six iterations; the sequential order bias takes five; and the nested structure bias takes four. This is because the feature vector bias learns a query that describes calls to \texttt{checkData}, \texttt{checkWidget}, and \texttt{error} regardless of their structure and order. In contrast, the sequential bias finds code that calls \texttt{checkWidget} before \texttt{error} and \texttt{getFont}. The nested structure bias finds code that calls \texttt{checkData} within an \texttt{if} statement, which respects the structure of the selected block. We find that a {\em flat} representation of code as a feature vector is not powerful enough to effectively capture the search intent in a few iterations. We need to incorporate the rich structural information already available in the code to achieve desired~performance. 
\vspace{-1mm}
\subsection{Labeling Positives and Negatives} \label{eval:bothlabel} 
To quantify the effect of using different types of labeled code examples, we assess {\tool} in three different conditions during query specialization---(1) incorporating both positive and negative examples, (2) considering negative examples only, and (3) considering positive examples only.

Figures~\ref{fig:onlyposnegpres} and~\ref{fig:onlyposnegpresrecall} compare the precision and recall in different conditions. When considering negative examples only, {\tool} achieves a high precision while converging in five iterations, whereas considering positives only takes more iterations to converge with lower precision. This is because {\tool} often starts with a general query in the first search iteration, resulting in a large number of search results that contain many false positives. Hence, it is essential to label several negatives to eliminate those false positives quickly. However, if we do not label any positives, we are likely to remove some true positives as well. This is reflected in the recall, where giving negative examples only has the least recall. Therefore, an optimal choice would be to use both positive and negative examples while iteratively specializing the query, which justifies the design of {\tool}.  

\subsection{\changemarker{Varying the Number of Annotated} Features}\label{eval:features}
Table~\ref{tab:featurevspr} summarizes the average precision and recall right after feature annotation, when varying the number of annotations from one to five. These features are randomly selected from the control structures, method calls and types in the seed example, and the results are averaged over ten runs. The result suggests that annotating more code elements in the seed example can more precisely express the search intent and thus increase the precision. However, the recall is negatively impacted by increasing the number of annotations. When the simulation experiment chooses more features, it is likely to choose features that are not shared across expected code fragments in the ground truth. Thus, the initial search query becomes too specific and therefore misses some expected code fragments. Let us consider the two similar but not identical code examples in Section~\ref{sec:motive} (Figures~\ref{fig:sel2} and~\ref{fig:posexample}). If a user tags \texttt{range==null\&\&i$<=$this.leadingPtr} (line 9 in Figure~\ref{fig:sel2}) as important, the recall decreases since the expected code example in Figure~\ref{fig:posexample} does not contain this feature. Though a real user can annotate any number of code elements, our experiment shows that tagging two code elements leads to the optimal precision and recall.

\begin{table}[t]
\centering
\caption{Varying \# of tagged features in the first iteration} 
\label{tab:featurevspr}
\begin{tabular}{@{}lrrrr@{}}
\toprule
          & \sf{1 Feature} & \sf{2 Features} & \sf{3 Features} &   \sf{4 Features} \\ \midrule
\textbf{Precision} & 0.16      & 0.47       & 0.68       & 0.80       \\ 
\textbf{Recall}    & 0.91      & 0.86       & 0.80       & 0.78       \\ 
\bottomrule
\end{tabular}
\end{table}

\begin{table}[t]
\centering
\caption{Varying \# of labeled examples in each iteration}
\label{tab:labelvspr}
\begin{tabular}{@{}lrrrr@{}}
\toprule
          & \sf{2 Labels} & \sf{3 Labels} & \sf{4 Labels}  & \sf{5 Labels}\\ \midrule
\textbf{Precision} &1.0     & 1.0     & 1.0     & 1.0\\ 
\textbf{Recall}    & 1.0     & 0.88     & 0.81     & 0.75\\ 
\textbf{\# Iterations} &7 &6 & 5&5 \\ 
\textbf{\# Total Labels}& 14 & 18 & 20 & 25 \\ 
\bottomrule
\end{tabular}
\vspace{-3mm}
\end{table}

\subsection{\changemarker{Varying the Number of Labeled Examples}}\label{eval:lables}
\begin{figure}[t]
      \includegraphics[width=\linewidth]{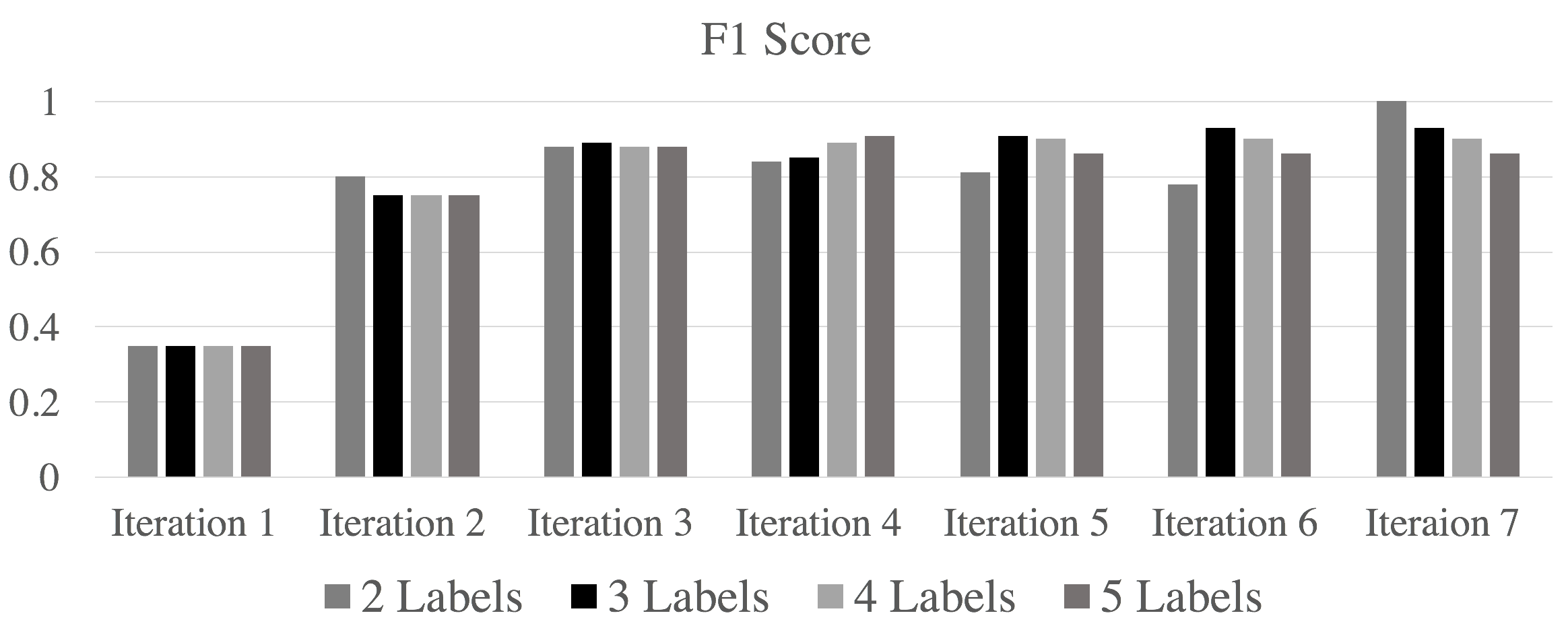}
  \caption{F1 score for different \# labels set at each iteration}
  \label{fig:labelacrossiter}
  \vspace{-3.5mm}
\end{figure}

The type and number of examples that a user labels in each iteration is an important aspect of active-learning based tool design. To quantify the benefit of iterative feedback over simultaneous labeling, we vary the number of labeled examples $n$ from two to five. Setting $n$ to four or five converges the fastest, with five iterations. When $n$ is three, {\tool} takes six iterations to converge. When $n$ is two, it takes seven. Table~\ref{tab:labelvspr} summarizes the average precision and recall for different number of labels in the final iteration. Overall, increasing the number of labeled examples maintains precision but does not necessarily increase recall. This is due to overfitting of labeled examples. From Table~\ref{tab:labelvspr} and Figure~\ref{fig:labelacrossiter}, $n=3$ gives a good trade-off between F1 score and the required amount of user feedback, and is more robust than $n=2$. 
Figure~\ref{fig:labelacrossiter} shows how user feedback helps, and why it is better to spread labeling effort over multiple iterations rather than providing all labeled examples at once. 

{\tool} takes an average of 20 seconds to parse a Java repository of about 20K methods and extract logic facts. This fact extraction is done only once in the beginning. {\tool} takes 6.8 seconds on average (median: 5 seconds) to infer a query and match the learned query to the entire codebase of about 670K facts. This realtime performance is appreciated by study participants, described in Section~\ref{sec:casestudy}.

\section{A Case Study with Users} 
\label{sec:casestudy}
To investigate how a user perceives the paradigm of annotating features and labeling examples during code search, we undertake a case study with three participants. The goal of this study is to investigate  how {\tool} may fit or augment current code search practices, rather than assessing how much benefit {\tool} provides in comparison to alternative code search approaches. Therefore, we use a case study research method~\cite{yin2008casestudy} to understand the in-depth context of how a user may \changemarker{interact with {\tool} to search desired code. In this study, a user tagged features, examined search results, and labeled them as positive and negative examples.} All participants were graduate students with multiple years of Java development experience. Participants reported that they use {\tt grep} or a regular expression search on a regular basis or have experience of using search features in IDEs. In terms of the study setup, we gave a tutorial about how to use {\tool}. This included reading a written tutorial followed by a step-by-step demonstration of tool features. The participants were given a seed example of {\ttt getSelectionText} from Eclipse SWT, which is part of Group 5 in Table~\ref{tab:dataset}. Since the participants were not familiar with the codebase, we provided two other similar snippets as a reference. Each session took about 30 minutes to 1 hour. Participants were encouraged to follow a {think-aloud} protocol to verbalize their thought process. 

\noindent\textbf{Do you think marking features in a snippet helps search?}
\begin{addmargin}[0.5em]{0.5em}
\textit{``I think marking features helps greatly when trying to find similar code snippets\textemdash as a human, my intuition is to identify key sections of code that I expect to be similar and ignore other sections that I expect to differ.''}
\newline
\textit{``It is obvious that when we are searching for a piece of code and we are expecting to have certain features in it, so marking features in a code snippet helps.''}
\end{addmargin}
Participants can easily identify important features in the seed example. Two of three participants annotated three features and one started with one feature. Participants found {\tool} useful for refactoring and reorganizing {\em fragmented yet similar} code snippets. They said that {\tool} would be more useful if integrated into code sharing websites such as GitHub and BitBucket by enabling advanced code search in large-scale repositories, because novice programmers would often prefer to look up functions with similar implementations for optimization opportunities.

\noindent\textbf{How do you think that labeling positive and negative examples fit in the search process?}
\begin{addmargin}[0.5em]{0.5em}
\textit{``As a developer, I will mentally do a quick evaluation of search results to determine if something is definitely negative (i.e., I can skip it) or potentially positive (i.e., it warrants more investigation). It's normally not an explicit process, but it makes sense to explicitly request feedback when interacting with tools.''}
\newline
\textit{``I really liked the iterative refinement of examples and the almost-realtime performance of the tool. One aspect that may be improved is, assigning a score or priority to the examples (based on how many features are satisfied) so that the user can prioritize which examples to inspect.''}
\end{addmargin}

\noindent\textbf{How do you think that an interactive and iterative approach such as ALICE compares to other search techniques that you used in the past?}
\begin{addmargin}[0.5em]{0.5em}
\textit{``ALICE seems useful for finding similar code snippets that span code blocks. It provides a functionality separate from other search techniques I've used such as IDE-based functionalities (e.g., call declaration/ method invocation locations) and standard {\tt grep} which is limited to single line searches. I think ALICE is generally less applicable, but also significantly more powerful. It also helps promote better understanding\textemdash for example, IDEs can show where a method is called from, but ALICE can easily show the context in which that method invocation appears.''}
\end{addmargin}
Participants said {\tool} was more powerful than standard search tools like {\ttt grep} and built-in search features in IDEs. 

\noindent\textbf{What do you like or not like about Alice?}
Overall, participants like the interactive feature, as it allows for refinement and builds on developer understanding. Some participants find the color scheme in {\tool} confusing due to the conflict with existing syntax highlighting in Eclipse. 
 
We observe that participants were able to easily recognize the important features in the code example and tag them in the first search iteration. Though participants had little experience with the codebase, they could still distinguish positive and negative examples without much effort. \changemarker{Table~\ref{tab:userlabeleffort} summarizes the number of examples returned in the first search iteration and the time taken for each user to refine the search. In particular, a user took an average of 35 seconds to inspect each example and categorize it as positive or negative. This indicates that the tool does not require much effort for a user to inspect and label examples.}
\begin{table}[t]
\centering
\caption{Time taken to label examples in Iteration 1}
\label{tab:userlabeleffort}
\begin{tabular}{@{}lrrrr@{}}
\toprule
          & \sf{\#Examples} & \sf{\#Positives} & \sf{\#Negatives}  & \sf{Time Taken(s)}\\ \midrule
\textbf{User 1} &8     & 1     & 1     & 20\\ 
\textbf{User 2}    & 437    & 0     & 2     & 55\\ 
\textbf{User 3} &8 &1 & 0&25\\ 
\bottomrule
\end{tabular}
\vspace{-3mm}
\end{table}

\section{Comparison} 
\label{sec:comparison} 
We compare {\tool} with an existing interactive code search technique called Critics. We choose Critics as a comparison baseline, since Critics also performs interactive search query refinement~\cite{zhang2015interactive}. {\tool} differs from Critics in two ways. First, in Critics, a user has to manually select code blocks such as an if-statement or a while loop and parameterizes contents, for example, by replacing {\ttt Foo f=new Foo()} to {\ttt \$T \$v=new \$T()}. Such interaction is time-consuming. To reduce the user burden, ALICE infers a syntactic pattern such as ``{\ttt an if-statement with a condition .*!=null inside a while loop}" from positive and negative methods. 
Second, Critics identifies similar code via tree matching, while {\tool} abstracts source code to logic facts and identifies similar code via ILP.

We run {\tool} on the public data set obtained from Critics's website.\footnote{https://github.com/tianyi-zhang/Critics} Table~\ref{tab:criticscomparison} summarizes the results of Critics vs.~{\tool}. In six out of seven cases, {\tool} achieves the same or better precision and recall with fewer iterations to converge, compared to Critics. In ID 4, {\tool} has low precision because the expected code contains a {\ttt switch} statement, which is currently not extracted by {\tool} as a logic fact. Extending current logic predicates to support more syntactic constructs remain as future work. 
\begin{table}[t]
\centering
\caption{Comparison Against Critics}
\resizebox{\linewidth}{!}{
\label{tab:criticscomparison}
\begin{tabular}{@{}p{0.6cm}rrr|rrr@{}}
\toprule
 &\multicolumn{3}{c}{\sf{Alice}} & \multicolumn{3}{c}{\sf{Critics}}\\
   \cmidrule{2-4}
   \cmidrule{5-7}
\sf{Critics ID} & \sf{Precision} & \sf{Recall} & \sf{Iterations} &\sf{Precision} & \sf{Recall} & \sf{Iterations} \\
\midrule
1 & 1.0 & 1.0 & 2 &1.0 &1.0 & 4\\
2 & 1.0 & 1.0 & 2 & 1.0 & 0.9 & 6 \\
3 & 1.0 & 1.0 & 1 & 1.0 & 0.88 & 6 \\
4 & 0.0 & 1.0 & 1 & 1.0 & 1.0 & 0 \\
5 & 1.0 & 1.0 & 3 & 1.0 & 1.0 & 7 \\
6 & 1.0 & 1.0 & 3 & 1.0 & 1.0 & 4 \\
7 & 1.0 & 1.0 & 3 & 1.0 & 0.33 & 3 \\
\midrule
\textbf{Average} & 0.86 & 1.0 & 2.1 & 1.0 & 0.87 & 4.3\\
\bottomrule
\end{tabular}
}%
\end{table}

\section{Discussion}
\noindent{\bf Noisy Oracle.} The simulation in Section~\ref{sec:evaluation} assumes that a user makes no mistakes when labeling examples. However, it is possible that a real user may label a positive example as negative, or even provide an inconsistent set of labels. We investigate how resilient {\tool} is to labeling mistakes and how quickly {\tool} can inform the user of such inconsistencies. We mutate our automated oracle with an error rate of 10\%, 20\%, and 40\%. Each of the 14 groups from the first dataset is run five times (70 trials) with different annotations, labels, and errors. In many cases, {\tool} reports inconsistencies in user labeling and provides immediate feedback on the infeasibility of specializing queries (33\% to 60\%). When {\tool} does not find any inconsistencies, {\tool} behaves robustly with respect to errors, eventually reaching 100\% precision. Table~\ref{tab:noisyoracle} summarizes the results.  

\begin{table}[t]
\centering
\caption{Sensitivity of {\tool} to labeling errors.} 
\label{tab:noisyoracle}
\begin{tabular}{@{}lrrrr@{}}
\toprule
&\multicolumn{3}{c}{\sf{Error Rates}} \\
\cmidrule{2-4}
&          \sf{ 10\%} & \sf{20\%} & \sf{40\%}\\ \midrule
\textbf{Precision} & 1.0      & 1.0       & 1.0             \\ 
\textbf{Recall}    & 0.95      & 0.90      & 0.93            \\ 
\textbf{\% Inconsistent Cases} & {33\%} & {60\%} & {54\%} \\

\bottomrule
\end{tabular}
\vspace{-3mm}
\end{table}

\vspace{1mm}
\noindent{\bf Threats to Validity.}
Regarding internal validity, the effectiveness of different inductive biases may depend on the extent and nature of code cloning in a codebase. For example, when there are many code clones with similar nested code structures ({\ttt while} and {\ttt if} statements), the nested structure may perform better than other inductive biases. The current simulation experiment is run on {\tool} by choosing one seed example from each group, by randomly selecting annotations from the selected seed, and by labeling a randomly chosen subset of returned results. To mitigate the impact of random selection, we repeat ten runs and report the average numbers. In terms of external validity, we assume that any user could easily annotate features and label examples. However, it is likely that a novice programmer might find it hard to identify important features. To mitigate this threat to validity, as future work, we will investigate the impact of different expertise levels.

\vspace{1mm}
\noindent{\bf Limitations and Future Work.}
Currently, we generate facts based on structural and intra-procedural control flow properties. Other types of analysis such as dataflow analysis or aliasing analysis could be used in identifying similar snippets. In addition, the query language itself can be extended to make it easier to capture the properties of desired code. For example, by introducing negations in the query language, a user can specify atoms that should not be included. There could be specializations that strictly require negations. However, in our experiments, empirically, we are always able to find a pattern without negations. As mentioned in Section~\ref{sec:approach}, our learning process is monotonic and to learn a different query, a user may need to start over. To overcome this, we may need backtracking and investigate new search algorithms that generalize and specialize the query in a different way.
\label{sec:discussion}

\section{Related Work}
\noindent{\bf Code Search and Clone Detection.} 
Different code search techniques and tools have been proposed to retrieve code examples from a large corpus of data~\cite{Holmes2005, sahavechaphan2006xsnippet, mcmillan2011portfolio, thummalapenta2007parseweb, brandt2010example, ponzanelli2014mining}. The most popular search approaches are based on text, regular expressions, constraints~\cite{stolee2014solving}, and natural language~\cite{bajracharya2006sourcerer,mcmillan2011portfolio,white2016deep,gu2018deep}. Exemplar~\cite{mcmillan2012exemplar} takes a natural language query as input and uses information retrieval and program analysis techniques to identify relevant code. Wang et al.~propose a dependence-based code search technique that matches a given pattern against system dependence graphs~\cite{wang2010matching}. XSnippet~\cite{sahavechaphan2006xsnippet} allows a user to search based on object instantiation using type hierarchy information from a given example. {\tool} differs from these search techniques in two ways. First, {\tool} allows a user to tag important features to construct an initial query. Second, {\tool} uses active learning to iteratively refine a query by leveraging positive vs.~negative labels. 

\changemarker{ALICE is fundamentally different from clone detectors~\cite{li2004cp,kamiya2002ccfinder,sajnani2016sourcerercc,deckard,roy2008nicad} in two ways. First, while clone detectors use a given internal representation such as a token string and a given similarity threshold to search for similar code, ALICE infers the commonality between positive examples, encodes them as a search template, and uses negative examples to decide what not to include in the template. Second, ALICE presents the search template as a logic query to a user, while clone detectors do not infer nor show a template to a user.}

\noindent{\bf Logic-Programming-Based Techniques.} JQuery is a code browsing tool\cite{janzen2003navigating} based on a logic query language. Users can interactively search by either typing a query in the UI or selecting a predefined template query. Hajiyev et al.~present an efficient and scalable code querying tool \cite{hajiyev2006codequest} that allows programmers to explore the relation between different parts of a codebase. A number of techniques use logic programming as an abstraction for detecting code smells \cite{wuyts1998declarative,tourwe2003identifying,gueheneuc2001using}. Many program analysis techniques abstract programs as logic facts and use Datalog rules, including pointer and call-graph analyses \cite{bravenboer2009strictly,smaragdakis2011pick}, concurrency analyses \cite{naik2006effective,naik2009effective}, datarace detection~\cite{mangal2015user}, security analyses \cite{martin2005finding}, etc. 
Apposcopy\cite{feng2014apposcopy} is a semantics-based Android malware detector, where a user provides a malware signature in Datalog. 
While {\tool} and this line of research both use logic programs as an underlying representation, {\tool} does not expect a user to know how to write a logic query nor requires having a set of pre-defined query templates. Instead, {\tool} allows the user to interactively and incrementally build a search query using active ILP.

\noindent{\bf Interactive Synthesis.} Some program synthesis techniques use input-output examples to infer a program and interactively refine its output~\cite{wang2017synthesizing,galenson2014codehint}. For instance, CodeHint~\cite{galenson2014codehint} is a dynamic code synthesis tool that uses runtime traces, a partial program sketch specification, and a probabilistic model to generate candidate expressions. Interactive disambiguation interfaces~\cite{li2014nalir, mayer2015user} aim to improve the accuracy of programming-by-example systems. {\tool} is similar to these in leveraging {\em interactivity}, but these do not target code search, do not use ILP, and do not assess the impact of iterative labeling and annotations.

\noindent{\bf Machine Learning.}
Active learning is often used when unlabeled data may be abundant or easy to come by, but training labels are difficult, time-consuming, or expensive to obtain~\cite{tong2001support,settles2012active}. 
An active learner may pose questions, usually in the form of unlabeled data instances to be labeled by an ``oracle'' (e.g., a human annotator). LOGAN-H is an ILP-based active learning approach~\cite{arias2007learning}. It learns clauses by either asking the oracle to label examples (membership queries) or to answer an equivalence query. 
Such oracles were first proposed by Angluin in the query-based learning formalism~\cite{angluin1988queries}.
Other approaches to inductive logic programming and relational learning are surveyed in~De Raedt~\cite{DBLP:ilpbook}.
\changemarker{Alrajeh et al.~integrate model checking and inductive learning to infer requirement specifications~\cite{alrajeh2013elaborating}.}
Other applications of ILP to software engineering include the work of Cohen~\cite{cohen1994recovering,cohen1995inductive}, to learn logical specifications from concrete program behavior.
Because ultimately our approach is not concerned with finding the right hypothesis, and only with retrieving the right code examples, it can also be thought of as a transductive learning problem~\cite{gammerman1998learning,chapelle2009semi}.

\label{sec:related}

\section{Conclusion}
{\tool} is the first approach that embodies the paradigm of active learning in the context of code search. Its algorithm is designed to leverage partial incremental feedback through tagging and labelling. {\tool} demonstrates realtime performance in constructing a new search query. Study participants resonate with {\tool}'s {\em interactive} approach and find it easy to describe a desired code pattern without much effort. 
Extensive simulation shows that leveraging both positive and negative labels together can help achieve high precision and recall. Tagging features is also necessary for minimizing the size of initial search space. Our experimental results justify the design choice of {\tool}, indicating that {\em interactivity} pays off\textemdash labeling a few in a spread out fashion is more effective than labeling many at a time. 
\section*{Acknowledgment}
Thanks to anonymous participants from the University of California, Los Angeles for their participation in the user study and to anonymous reviewers for their valuable feedback. This work is supported by NSF grants CCF-1764077, CCF-1527923, CCF-1460325, CCF-1837129, CCF-1723773, IIS-1633857, ONR grant N00014-18-1-2037, DARPA grant N66001-17-2-4032 and an Intel CAPA grant.

\newpage
\bibliographystyle{IEEEtran}
\bibliography{draft}

\begin{thebibliography}{10}
\providecommand{\url}[1]{#1}
\csname url@samestyle\endcsname
\providecommand{\newblock}{\relax}
\providecommand{\bibinfo}[2]{#2}
\providecommand{\BIBentrySTDinterwordspacing}{\spaceskip=0pt\relax}
\providecommand{\BIBentryALTinterwordstretchfactor}{4}
\providecommand{\BIBentryALTinterwordspacing}{\spaceskip=\fontdimen2\font plus
\BIBentryALTinterwordstretchfactor\fontdimen3\font minus
  \fontdimen4\font\relax}
\providecommand{\BIBforeignlanguage}[2]{{%
\expandafter\ifx\csname l@#1\endcsname\relax
\typeout{** WARNING: IEEEtran.bst: No hyphenation pattern has been}%
\typeout{** loaded for the language `#1'. Using the pattern for}%
\typeout{** the default language instead.}%
\else
\language=\csname l@#1\endcsname
\fi
#2}}
\providecommand{\BIBdecl}{\relax}
\BIBdecl

\bibitem{kim2006memories}
S.~Kim, K.~Pan, and E.~Whitehead~Jr, ``Memories of bug fixes,'' in
  \emph{Proceedings of the 14th ACM SIGSOFT international symposium on
  Foundations of software engineering}.\hskip 1em plus 0.5em minus 0.4em\relax
  ACM, 2006, pp. 35--45.

\bibitem{nguyen2010recurring}
T.~T. Nguyen, H.~A. Nguyen, N.~H. Pham, J.~Al-Kofahi, and T.~N. Nguyen,
  ``Recurring bug fixes in object-oriented programs,'' in \emph{Software
  Engineering, 2010 ACM/IEEE 32nd International Conference on}, vol.~1.\hskip
  1em plus 0.5em minus 0.4em\relax IEEE, 2010, pp. 315--324.

\bibitem{park2012empirical}
J.~Park, M.~Kim, B.~Ray, and D.-H. Bae, ``An empirical study of supplementary
  bug fixes,'' in \emph{Proceedings of the 9th IEEE Working Conference on
  Mining Software Repositories}.\hskip 1em plus 0.5em minus 0.4em\relax IEEE
  Press, 2012, pp. 40--49.

\bibitem{cordy2006txl}
J.~R. Cordy, ``The txl source transformation language,'' \emph{Science of
  Computer Programming}, vol.~61, no.~3, pp. 190--210, 2006.

\bibitem{wang2010matching}
X.~Wang, D.~Lo, J.~Cheng, L.~Zhang, H.~Mei, and J.~X. Yu, ``Matching
  dependence-related queries in the system dependence graph,'' in
  \emph{Proceedings of the IEEE/ACM international conference on Automated
  software engineering}.\hskip 1em plus 0.5em minus 0.4em\relax ACM, 2010, pp.
  457--466.

\bibitem{andersen2010generic}
J.~Andersen and J.~L. Lawall, ``Generic patch inference,'' \emph{Automated
  software engineering}, vol.~17, no.~2, pp. 119--148, 2010.

\bibitem{pham2010detection}
N.~H. Pham, T.~T. Nguyen, H.~A. Nguyen, and T.~N. Nguyen, ``Detection of
  recurring software vulnerabilities,'' in \emph{Proceedings of the IEEE/ACM
  international conference on Automated software engineering}.\hskip 1em plus
  0.5em minus 0.4em\relax ACM, 2010, pp. 447--456.

\bibitem{meng2011systematic}
N.~Meng, M.~Kim, and K.~S. McKinley, ``Systematic editing: generating program
  transformations from an example,'' \emph{ACM SIGPLAN Notices}, vol.~46,
  no.~6, pp. 329--342, 2011.

\bibitem{meng2013lase}
------, ``Lase: locating and applying systematic edits by learning from
  examples,'' in \emph{Proceedings of the 2013 International Conference on
  Software Engineering}.\hskip 1em plus 0.5em minus 0.4em\relax IEEE Press,
  2013, pp. 502--511.

\bibitem{rolim2017learning}
R.~Rolim, G.~Soares, L.~D'Antoni, O.~Polozov, S.~Gulwani, R.~Gheyi, R.~Suzuki,
  and B.~Hartmann, ``Learning syntactic program transformations from
  examples,'' in \emph{Proceedings of the 39th International Conference on
  Software Engineering}.\hskip 1em plus 0.5em minus 0.4em\relax IEEE Press,
  2017, pp. 404--415.

\bibitem{zhang2015interactive}
T.~Zhang, M.~Song, J.~Pinedo, and M.~Kim, ``Interactive code review for
  systematic changes,'' in \emph{Proceedings of the 37th International
  Conference on Software Engineering-Volume 1}.\hskip 1em plus 0.5em minus
  0.4em\relax IEEE Press, 2015, pp. 111--122.

\bibitem{muggleton1994inductive}
S.~Muggleton and L.~De~Raedt, ``Inductive logic programming: Theory and
  methods,'' \emph{The Journal of Logic Programming}, vol.~19, pp. 629--679,
  1994.

\bibitem{DBLP:ilpbook}
\BIBentryALTinterwordspacing
L.~D. Raedt, \emph{Logical and relational learning}, ser. Cognitive
  Technologies.\hskip 1em plus 0.5em minus 0.4em\relax Springer, 2008.
  [Online]. Available: \url{https://doi.org/10.1007/978-3-540-68856-3}
\BIBentrySTDinterwordspacing

\bibitem{neuralnetworks}
S.~Haykin, \emph{Neural Networks: A Comprehensive Foundation (3rd
  Edition)}.\hskip 1em plus 0.5em minus 0.4em\relax Upper Saddle River, NJ,
  USA: Prentice-Hall, Inc., 2007.

\bibitem{muggleton1990efficient}
S.~Muggleton, C.~Feng \emph{et~al.}, \emph{Efficient induction of logic
  programs}.\hskip 1em plus 0.5em minus 0.4em\relax Citeseer, 1990.

\bibitem{cohen1994recovering}
W.~W. Cohen, ``Recovering software specifications with inductive logic
  programming,'' in \emph{AAAI}, vol.~94, 1994, pp. 1--4.

\bibitem{bratko1993inductive}
I.~Bratko and M.~Grobelnik, ``Inductive learning applied to program
  construction and verification,'' 1993.

\bibitem{alrajeh2013elaborating}
D.~Alrajeh, J.~Kramer, A.~Russo, and S.~Uchitel, ``Elaborating requirements
  using model checking and inductive learning,'' \emph{IEEE Transactions on
  Software Engineering}, vol.~39, no.~3, pp. 361--383, 2013.

\bibitem{starke2009working}
J.~Starke, C.~Luce, and J.~Sillito, ``Working with search results,'' in
  \emph{Search-Driven Development-Users, Infrastructure, Tools and Evaluation,
  2009. SUITE'09. ICSE Workshop on}.\hskip 1em plus 0.5em minus 0.4em\relax
  IEEE, 2009, pp. 53--56.

\bibitem{ahmad2018automatically}
M.~B.~S. Ahmad and A.~Cheung, ``Automatically leveraging mapreduce frameworks
  for data-intensive applications,'' in \emph{Proceedings of the 2018
  International Conference on Management of Data}.\hskip 1em plus 0.5em minus
  0.4em\relax ACM, 2018, pp. 1205--1220.

\bibitem{ahmad2016leveraging}
------, ``Leveraging parallel data processing frameworks with verified
  lifting,'' \emph{arXiv preprint arXiv:1611.07623}, 2016.

\bibitem{brandt2009two}
J.~Brandt, P.~J. Guo, J.~Lewenstein, M.~Dontcheva, and S.~R. Klemmer, ``Two
  studies of opportunistic programming: interleaving web foraging, learning,
  and writing code,'' in \emph{Proceedings of the SIGCHI Conference on Human
  Factors in Computing Systems}.\hskip 1em plus 0.5em minus 0.4em\relax ACM,
  2009, pp. 1589--1598.

\bibitem{duala2012asking}
E.~Duala-Ekoko and M.~P. Robillard, ``Asking and answering questions about
  unfamiliar apis: An exploratory study,'' in \emph{Proceedings of the 34th
  International Conference on Software Engineering}.\hskip 1em plus 0.5em minus
  0.4em\relax IEEE Press, 2012, pp. 266--276.

\bibitem{settles2012active}
B.~Settles, ``Active learning,'' \emph{Synthesis Lectures on Artificial
  Intelligence and Machine Learning}, vol.~6, no.~1, pp. 1--114, 2012.

\bibitem{de2008logical}
L.~De~Raedt, \emph{Logical and relational learning}.\hskip 1em plus 0.5em minus
  0.4em\relax Springer Science \& Business Media, 2008.

\bibitem{flach1994simply}
P.~A. Flach, ``Simply logical intelligent reasoning by example,'' 1994.

\bibitem{costa2012yap}
V.~S. Costa, R.~Rocha, and L.~Damas, ``The yap prolog system,'' \emph{Theory
  and Practice of Logic Programming}, vol.~12, no. 1-2, pp. 5--34, 2012.

\bibitem{yin2008casestudy}
R.~K. Yin, \emph{Case Study Research: Design and Methods (Applied Social
  Research Methods)}, fourth edition.~ed.\hskip 1em plus 0.5em minus
  0.4em\relax Sage Publications, 2008.

\bibitem{Holmes2005}
R.~Holmes and G.~C. Murphy, ``Using structural context to recommend source code
  examples,'' in \emph{ICSE '05: Proceedings of the 27th International
  Conference on Software Engineering}.\hskip 1em plus 0.5em minus 0.4em\relax
  New York, NY, USA: ACM Press, 2005, pp. 117--125.

\bibitem{sahavechaphan2006xsnippet}
N.~Sahavechaphan and K.~Claypool, ``Xsnippet: Mining for sample code,''
  \emph{ACM Sigplan Notices}, vol.~41, no.~10, pp. 413--430, 2006.

\bibitem{mcmillan2011portfolio}
C.~McMillan, M.~Grechanik, D.~Poshyvanyk, Q.~Xie, and C.~Fu, ``Portfolio:
  finding relevant functions and their usage,'' in \emph{Proceedings of the
  33rd International Conference on Software Engineering}.\hskip 1em plus 0.5em
  minus 0.4em\relax ACM, 2011, pp. 111--120.

\bibitem{thummalapenta2007parseweb}
S.~Thummalapenta and T.~Xie, ``Parseweb: a programmer assistant for reusing
  open source code on the web,'' in \emph{Proceedings of the twenty-second
  IEEE/ACM international conference on Automated software engineering}.\hskip
  1em plus 0.5em minus 0.4em\relax ACM, 2007, pp. 204--213.

\bibitem{brandt2010example}
J.~Brandt, M.~Dontcheva, M.~Weskamp, and S.~R. Klemmer, ``Example-centric
  programming: integrating web search into the development environment,'' in
  \emph{Proceedings of the SIGCHI Conference on Human Factors in Computing
  Systems}.\hskip 1em plus 0.5em minus 0.4em\relax ACM, 2010, pp. 513--522.

\bibitem{ponzanelli2014mining}
L.~Ponzanelli, G.~Bavota, M.~Di~Penta, R.~Oliveto, and M.~Lanza, ``Mining
  stackoverflow to turn the ide into a self-confident programming prompter,''
  in \emph{Proceedings of the 11th Working Conference on Mining Software
  Repositories}.\hskip 1em plus 0.5em minus 0.4em\relax ACM, 2014, pp.
  102--111.

\bibitem{stolee2014solving}
K.~T. Stolee, S.~Elbaum, and D.~Dobos, ``Solving the search for source code,''
  \emph{ACM Transactions on Software Engineering and Methodology (TOSEM)},
  vol.~23, no.~3, p.~26, 2014.

\bibitem{bajracharya2006sourcerer}
S.~Bajracharya, T.~Ngo, E.~Linstead, Y.~Dou, P.~Rigor, P.~Baldi, and C.~Lopes,
  ``Sourcerer: a search engine for open source code supporting structure-based
  search,'' in \emph{Companion to the 21st ACM SIGPLAN symposium on
  Object-oriented programming systems, languages, and applications}.\hskip 1em
  plus 0.5em minus 0.4em\relax ACM, 2006, pp. 681--682.

\bibitem{white2016deep}
M.~White, M.~Tufano, C.~Vendome, and D.~Poshyvanyk, ``Deep learning code
  fragments for code clone detection,'' in \emph{Proceedings of the 31st
  IEEE/ACM International Conference on Automated Software Engineering}.\hskip
  1em plus 0.5em minus 0.4em\relax ACM, 2016, pp. 87--98.

\bibitem{gu2018deep}
X.~Gu, H.~Zhang, and S.~Kim, ``Deep code search,'' in \emph{Proceedings of the
  40th International Conference on Software Engineering}.\hskip 1em plus 0.5em
  minus 0.4em\relax ACM, 2018, pp. 933--944.

\bibitem{mcmillan2012exemplar}
C.~McMillan, M.~Grechanik, D.~Poshyvanyk, C.~Fu, and Q.~Xie, ``Exemplar: A
  source code search engine for finding highly relevant applications,''
  \emph{IEEE Transactions on Software Engineering}, vol.~38, no.~5, pp.
  1069--1087, 2012.

\bibitem{li2004cp}
Z.~Li, S.~Lu, S.~Myagmar, and Y.~Zhou, ``Cp-miner: A tool for finding
  copy-paste and related bugs in operating system code.'' in \emph{OSdi},
  vol.~4, no.~19, 2004, pp. 289--302.

\bibitem{kamiya2002ccfinder}
T.~Kamiya, S.~Kusumoto, and K.~Inoue, ``Ccfinder: a multilinguistic token-based
  code clone detection system for large scale source code,'' \emph{IEEE
  Transactions on Software Engineering}, vol.~28, no.~7, pp. 654--670, 2002.

\bibitem{sajnani2016sourcerercc}
H.~Sajnani, V.~Saini, J.~Svajlenko, C.~K. Roy, and C.~V. Lopes, ``Sourcerercc:
  Scaling code clone detection to big-code,'' in \emph{Software Engineering
  (ICSE), 2016 IEEE/ACM 38th International Conference on}.\hskip 1em plus 0.5em
  minus 0.4em\relax IEEE, 2016, pp. 1157--1168.

\bibitem{deckard}
\BIBentryALTinterwordspacing
L.~Jiang, G.~Misherghi, Z.~Su, and S.~Glondu, ``Deckard: Scalable and accurate
  tree-based detection of code clones,'' in \emph{Proceedings of the 29th
  International Conference on Software Engineering}, ser. ICSE '07.\hskip 1em
  plus 0.5em minus 0.4em\relax Washington, DC, USA: IEEE Computer Society,
  2007, pp. 96--105. [Online]. Available:
  \url{http://dx.doi.org/10.1109/ICSE.2007.30}
\BIBentrySTDinterwordspacing

\bibitem{roy2008nicad}
C.~K. Roy and J.~R. Cordy, ``Nicad: Accurate detection of near-miss intentional
  clones using flexible pretty-printing and code normalization,'' in
  \emph{Program Comprehension, 2008. ICPC 2008. The 16th IEEE International
  Conference on}.\hskip 1em plus 0.5em minus 0.4em\relax IEEE, 2008, pp.
  172--181.

\bibitem{janzen2003navigating}
D.~Janzen and K.~De~Volder, ``Navigating and querying code without getting
  lost,'' in \emph{Proceedings of the 2nd international conference on
  Aspect-oriented software development}.\hskip 1em plus 0.5em minus 0.4em\relax
  ACM, 2003, pp. 178--187.

\bibitem{hajiyev2006codequest}
E.~Hajiyev, M.~Verbaere, and O.~De~Moor, ``Codequest: Scalable source code
  queries with datalog,'' in \emph{European Conference on Object-oriented
  Programming}.\hskip 1em plus 0.5em minus 0.4em\relax Springer, 2006, pp.
  2--27.

\bibitem{wuyts1998declarative}
R.~Wuyts, ``Declarative reasoning about the structure of object-oriented
  systems,'' in \emph{Technology of Object-Oriented Languages, 1998. TOOLS 26.
  Proceedings}.\hskip 1em plus 0.5em minus 0.4em\relax IEEE, 1998, pp.
  112--124.

\bibitem{tourwe2003identifying}
T.~Tourw{\'e} and T.~Mens, ``Identifying refactoring opportunities using logic
  meta programming,'' in \emph{Software Maintenance and Reengineering, 2003.
  Proceedings. Seventh European Conference on}.\hskip 1em plus 0.5em minus
  0.4em\relax IEEE, 2003, pp. 91--100.

\bibitem{gueheneuc2001using}
Y.-G. Gu{\'e}h{\'e}neuc and H.~Albin-Amiot, ``Using design patterns and
  constraints to automate the detection and correction of inter-class design
  defects,'' in \emph{Technology of Object-Oriented Languages and Systems,
  2001. TOOLS 39. 39th International Conference and Exhibition on}.\hskip 1em
  plus 0.5em minus 0.4em\relax IEEE, 2001, pp. 296--305.

\bibitem{bravenboer2009strictly}
M.~Bravenboer and Y.~Smaragdakis, ``Strictly declarative specification of
  sophisticated points-to analyses,'' \emph{ACM SIGPLAN Notices}, vol.~44,
  no.~10, pp. 243--262, 2009.

\bibitem{smaragdakis2011pick}
Y.~Smaragdakis, M.~Bravenboer, and O.~Lhot{\'a}k, ``Pick your contexts well:
  understanding object-sensitivity,'' in \emph{ACM SIGPLAN Notices}, vol.~46,
  no.~1.\hskip 1em plus 0.5em minus 0.4em\relax ACM, 2011, pp. 17--30.

\bibitem{naik2006effective}
M.~Naik, A.~Aiken, and J.~Whaley, \emph{Effective static race detection for
  Java}.\hskip 1em plus 0.5em minus 0.4em\relax ACM, 2006, vol.~41, no.~6.

\bibitem{naik2009effective}
M.~Naik, C.-S. Park, K.~Sen, and D.~Gay, ``Effective static deadlock
  detection,'' in \emph{Software Engineering, 2009. ICSE 2009. IEEE 31st
  International Conference on}.\hskip 1em plus 0.5em minus 0.4em\relax IEEE,
  2009, pp. 386--396.

\bibitem{mangal2015user}
R.~Mangal, X.~Zhang, A.~V. Nori, and M.~Naik, ``A user-guided approach to
  program analysis,'' in \emph{Proceedings of the 2015 10th Joint Meeting on
  Foundations of Software Engineering}.\hskip 1em plus 0.5em minus 0.4em\relax
  ACM, 2015, pp. 462--473.

\bibitem{martin2005finding}
M.~Martin, B.~Livshits, and M.~S. Lam, ``Finding application errors and
  security flaws using pql: a program query language,'' \emph{ACM SIGPLAN
  Notices}, vol.~40, no.~10, pp. 365--383, 2005.

\bibitem{feng2014apposcopy}
Y.~Feng, S.~Anand, I.~Dillig, and A.~Aiken, ``Apposcopy: Semantics-based
  detection of android malware through static analysis,'' in \emph{Proceedings
  of the 22nd ACM SIGSOFT International Symposium on Foundations of Software
  Engineering}.\hskip 1em plus 0.5em minus 0.4em\relax ACM, 2014, pp. 576--587.

\bibitem{wang2017synthesizing}
C.~Wang, A.~Cheung, and R.~Bodik, ``Synthesizing highly expressive sql queries
  from input-output examples,'' in \emph{ACM SIGPLAN Notices}, vol.~52,
  no.~6.\hskip 1em plus 0.5em minus 0.4em\relax ACM, 2017, pp. 452--466.

\bibitem{galenson2014codehint}
J.~Galenson, P.~Reames, R.~Bodik, B.~Hartmann, and K.~Sen, ``Codehint: Dynamic
  and interactive synthesis of code snippets,'' in \emph{Proceedings of the
  36th International Conference on Software Engineering}.\hskip 1em plus 0.5em
  minus 0.4em\relax ACM, 2014, pp. 653--663.

\bibitem{li2014nalir}
F.~Li and H.~V. Jagadish, ``Nalir: an interactive natural language interface
  for querying relational databases,'' in \emph{Proceedings of the 2014 ACM
  SIGMOD international conference on Management of data}.\hskip 1em plus 0.5em
  minus 0.4em\relax ACM, 2014, pp. 709--712.

\bibitem{mayer2015user}
M.~Mayer, G.~Soares, M.~Grechkin, V.~Le, M.~Marron, O.~Polozov, R.~Singh,
  B.~Zorn, and S.~Gulwani, ``User interaction models for disambiguation in
  programming by example,'' in \emph{Proceedings of the 28th Annual ACM
  Symposium on User Interface Software \& Technology}.\hskip 1em plus 0.5em
  minus 0.4em\relax ACM, 2015, pp. 291--301.

\bibitem{tong2001support}
S.~Tong and E.~Chang, ``Support vector machine active learning for image
  retrieval,'' in \emph{Proceedings of the ninth ACM international conference
  on Multimedia}.\hskip 1em plus 0.5em minus 0.4em\relax ACM, 2001, pp.
  107--118.

\bibitem{arias2007learning}
M.~Arias, R.~Khardon, and J.~Maloberti, ``Learning horn expressions with
  logan-h,'' \emph{Journal of Machine Learning Research}, vol.~8, no. Mar, pp.
  549--587, 2007.

\bibitem{angluin1988queries}
D.~Angluin, ``Queries and concept learning,'' \emph{Machine learning}, vol.~2,
  no.~4, pp. 319--342, 1988.

\bibitem{cohen1995inductive}
W.~W. Cohen, ``Inductive specification recovery: Understanding software by
  learning from example behaviors,'' \emph{Automated Software Engineering},
  vol.~2, no.~2, pp. 107--129, 1995.

\bibitem{gammerman1998learning}
A.~Gammerman, V.~Vovk, and V.~Vapnik, ``Learning by transduction,'' in
  \emph{Proceedings of the Fourteenth conference on Uncertainty in artificial
  intelligence}.\hskip 1em plus 0.5em minus 0.4em\relax Morgan Kaufmann
  Publishers Inc., 1998, pp. 148--155.

\bibitem{chapelle2009semi}
O.~Chapelle, B.~Scholkopf, and A.~Zien, ``Semi-supervised learning,''
  \emph{IEEE Transactions on Neural Networks}, vol.~20, no.~3, pp. 542--542,
  2009.

\end{thebibliography}

\end{document}